\documentstyle[amssymb,graphicx,prd,aps,preprint,eqsecnum]{revtex} % RevTex
\tightenlines
\def\mtrm#1{{\rm #1}}             % RevTex
\def\mtca#1{{\cal #1}}            % RevTex
\def\dirac{\llap{/\,}}
\def\lcr{\left\langle\!\left\langle}
\def\rcr{\right\rangle\!\right\rangle}
\def\nunf{_\nu^{(N_f)}}
\def\nflav{^{(N_f)}}
\def\ncrit{n_\mtrm{crit}(N_c)}
\def\lo{leading order}
\def\nlo{next-to-leading order}
\def\leff{\mtca{L}_\mtrm{eff}}

\begin{document}

\preprint{IPNO/DR 99-32}

\title{Finite-volume analysis of\\
$N_f$-induced chiral phase transitions
\thanks{Work partially supported by the EEC, TMR-CT98-0169, EURODAPHNE
Network.}}
\author{S. Descotes and J. Stern}
\address{IPN,
Groupe de Physique Th\'eorique, Universit\'e de Paris-Sud,\\
F-91406 Orsay Cedex, France}

\maketitle                     % RevTex

\begin{abstract}
In the framework of Euclidean QCD on a torus, we study the spectrum of the
Dirac operator through inverse moments of its eigenvalues, averaged over 
topological sets of gluonic configurations. The large-volume
dependence of these sums is related to chiral order parameters. We sketch
how these results may be applied to lattice simulations in order to
investigate the chiral phase transitions occurring when $N_f$ increases. In
particular, we demonstrate how Dirac inverse moments at
different volumes could be compared to detect in a clean way 
the phase transition triggered by the suppression of the quark condensate
and by the enhancement of the Zweig-rule violation in the vacuum channel.
\end{abstract}

\vskip1cm

Pacs numbers: 11.30.Rd, 12.39.Fe, 12.38.Gc, 02.70.Fj.

\newpage

\tableofcontents

\newpage

\section{Introduction} \label{introduction}

Understanding the Spontaneous Breakdown of Chiral Symmetry (SB$\chi$S) remains one of the
most challenging non-perturbative problems of QCD. Forthcoming experiments
\cite{daphne,mainz,exper} should reveal some of
its features, at least in the non-strange sector in which the effective
number of light quark flavours is minimal ($N_f=2$). It is generally
expected that if $N_f$ increases (keeping the number of colours $N_c$
fixed), the theory meets phase transitions and the chiral symmetry is
eventually restored. The argument is originally based on properties of
the QCD $\beta$-function in perturbation theory. The well-known
statement of the ``end of asymptotic freedom'' for $N_f\geq 11N_c/2$
\cite{betafunc} has been further completed by the analysis of the so-called
``conformal window'' \cite{Banks-Zaks} suggesting a restoration of chiral
symmetry for lower $N_f$, such as $N_f\sim 10$ (for $N_c=3$) \cite{Grunberg}.
Less perturbative and more model-dependent investigations, based on a gap
equation \cite{gapeq} or on a ``liquid instanton model'' \cite{instmodel}, 
also indicate that a chiral phase transition could occur for $N_f$ substantially
below $11N_c/2$.

It is important to understand, at least qualitatively, the non-perturbative
origin of the suppression of chiral order parameters for an increasing
$N_f$. We have recently argued \cite{Descotes-Girlanda-Stern} that such a
suppression might result from a paramagnetic effect of light (massless) quark
loops \cite{paramag}, i.e. it could be due to ``sea quarks'' and, consequently, it could
escape a detection in quenched lattice simulations, or in any other
approach neglecting the fermion determinant. An
appropriate framework to develop these ideas and to ask precise questions is the
formulation of QCD in an Euclidean box $L\times L \times L \times L$, 
with periodic (antiperiodic) boundary conditions for gluon
(fermion) fields, up to a gauge transformation. In this framework,
the SB$\chi$S pattern is reflected by the dynamics of lowest eigenvalues
of the Dirac operator:
\begin{equation}
H[G]=\gamma_\mu(\partial_\mu+iG_\mu). \label{diracoper}
\end{equation}
This Hermitian operator has a symmetric spectrum with respect to zero:
$\{H,\gamma_5\}=0$. Positive eigenvalues $\lambda_n$ are labeled in
ascending order by a positive integer (one further denotes
$\lambda_{-n}=-\lambda_n$ and  $\phi_{-n}=\gamma_5\phi_n$ for the
corresponding eigenvectors).
SB$\chi$S is related to a particularly dense accumulation of eigenvalues
around zero \cite{Vafa-Witten,Banks-Casher,Stern,Leut-Smil}. Models of such an
accumulation in terms of random matrices \cite{randmat} or instantons
\cite{instantons} have been proposed. Some chiral order parameters are
entirely dominated by the infrared extremity of the spectrum of the Dirac
operator (\ref{diracoper}). This makes them particularly sensitive to the
statistical weight given to smallest Dirac eigenvalues in the functional
integral, which is suppressed in the massless limit by the $N_f$-th power of
the fermion determinant. A good example is the quark condensate, defined by:
\begin{equation}
\Sigma(N_f)=-\lim_{m_1,m_2,\ldots m_{N_f}\to 0} \langle 0|\bar{u}u|0\rangle,
  \label{defcondqrk}
\end{equation}
where $m_1$\ldots $m_{N_f}$ denote the $N_f$ lightest quark masses and $u$
represents the lightest quark field. $\Sigma(N_f)$ receives exclusive 
contributions
from the smallest Dirac eigenvalues that behave in average as $1/L^4$, and 
it is consequently expected to be the most sensitive order parameter 
to the variation of
$N_f$ and to a phase transition. Other order parameters are less sensitive,
like $F^2(N_f)$, defined as the $\mtrm{SU}_L(N_f)\times\mtrm{SU}_R(N_f)$ limit of
the coupling of the Goldstone bosons to the axial current:
\begin{equation}
F^2(N_f)=\lim_{m_1,m_2,\ldots m_{N_f}\to 0} F^2_\pi. \label{defdecons}
\end{equation}
$F^2(N_f)$ may be non-zero due to Dirac eigenvalues accumulating as $1/L^2$
\cite{Stern}. For this reason, $F^2(N_f)$ should exhibit a weaker
$N_f$-dependence than $\Sigma(N_f)$. Finally, observables with no particular
sensitivity to the infrared edge of the Dirac spectrum ($\rho$-mass, string
tension, etc\ldots) have no reason to be strongly affected by the fermion
determinant and by the $N_f$-dependence.

Let us first consider the thermodynamical limit and denote by $\ncrit$ the
critical value of $N_f$ at which the first chiral phase transition takes
place. Just below $\ncrit$, the order parameter $\Sigma(N_f)$ drops out,
whereas its fluctuations may be expected to become important. We have shown
\cite{Descotes-Girlanda-Stern} that the latter would manifest itself by an
enhancement of the Zweig-rule violation just in the vacuum channel
$J^{PC}=0^{++}$. An important Zweig-rule violation is precisely observed in
the scalar channel \cite{scalar}, and nowhere else (with the exception of the pseudoscalar
channel driven by the axial anomaly). Whilst the signature of a nearby phase
transition is rather clear just below $\ncrit$, it is more speculative and
ambiguous above the critical point. First, above $\ncrit$, colour might
still be confined (confinement has no obvious relation to small Dirac
eigenvalues). Second, despite $\Sigma(N_f)=0$, the chiral symmetry need not
be completely restored. The Goldstone bosons coupling to conserved axial
currents with the strength $F(N_f)$ might survive to the $N_f$-induced phase
transition. This is reflected by the possibility that the $N_f$-sensitivity
and suppression of the order parameter $F^2(N_f)$ might be considerably
weaker than in the case of the quark condensate \cite{Stern}. Of course,
this is a highly non-trivial possibility, which presumably depends on the
existence of a non-perturbative fixed point in the renormalization group
flow\footnote{If one sticks to cut-off-dependent bare quantities, it is
possible to argue that $\Sigma=0$ would imply $F=0$, i.e. the full symmetry
restoration \cite{Shifman}. This argument is however based on an inequality
for which it is by no means obvious that it survives in the full
renormalized theory.}. Here, we take as a working hypothesis that above
$\ncrit$, a partial SB$\chi$S still occurs, due to $F^2(N_f)\neq 0$. The results
of our paper allow, in particular, to test this hypothesis.

The central question remains how far is $\ncrit$ (for $N_c=3$) from the
real world, in which the number of light quarks hardly exceeds $N_f=2-3$.
Some recent investigations actually indicate that $n_\mtrm{crit}(3)$ could
be rather small, and/or that the real world could already feel the influence
of a nearby phase transition. First, some lattice simulations with
dynamical fermions observe a strong $N_f$-dependence of SB$\chi$S signals
for $N_f$ as low as 4-6 \cite{lattflav1,lattflav2}. Second, a method based
on a well-convergent chiral sum rule has been proposed, which allows to
study phenomenologically the variation of $\Sigma(N_f)$ for small $N_f$
\cite{Moussallam}. It has been found that existing experimental information
on the Zweig-rule violation in the scalar channel leads to a large reduction
of $\Sigma(N_f)$ already between $N_f=2$ and $N_f=3$.

The purpose of this paper is to analyze in a model-independent way how
$N_f$-induced chiral phase transitions manifest themselves in the
finite-volume partition function. In particular, we shall investigate the
volume dependence of the inverse spectral moments of the Dirac operator
(\ref{diracoper}):
\begin{equation}
\sigma_k=\sum_{n>0} \frac{1}{(\lambda_n[G])^k}, \label{invmom}
\end{equation}
averaged over topological sets of gluonic configurations. For
$N_f\ll \ncrit$, the leading large-volume behaviour of such inverse moments
has been worked out in details by Leutwyler and Smilga \cite{Leut-Smil}. In
order to investigate how this result is modified in the vicinity and above
$\ncrit$, we rely on the basic observations and methods of
Ref.~\cite{Leut-Smil}.
For large sizes of the box ($\Lambda_H L\gg 1$ with $\Lambda_H\sim 1$ Gev),
heavy excitations are exponentially suppressed in the
partition function, which is then dominated by the lightest states, the
pseudo-Goldstone bosons of SB$\chi$S. This leads to an effective description
in terms of the Chiral Perturbation Theory ($\chi$PT) \cite{schipt,gchipt}, 
and it can be matched with QCD, yielding the desired information concerning
the infrared properties of the Dirac spectrum. Moreover, the effective
Lagrangian is identical to its infinite-volume counterpart, provided that periodic
boundary conditions are used \cite{Gass-Leut2}. 

If $N_f$ lies far below $\ncrit$, the quark condensate is large
and $\sigma_k$ behaves at large (but finite) volumes according to the
asymptotic behaviour derived by Leutwyler and Smilga \cite{Leut-Smil},
using Standard $\chi$PT \cite{schipt}.
Above $\ncrit$, the quark condensate vanishes,
and the previous analysis cannot be applied. However, if chiral symmetry
is still partially broken, the matching with $\chi$PT remains possible and it
leads to a clear-cut change in the large-volume behaviour of $\sigma_k$ :
expressed through their inverse moments,
the average behaviour of the lowest eigenvalues for $L\to\infty$ should
turn from $1/L^4$ ($\langle\bar{q}q\rangle\neq 0$) into $1/L^2$
$(\langle\bar{q}q\rangle=0)$ \cite{Stern}. When we approach the critical
point with $N_f$ near but under $\ncrit$, significant
discrepancies from the asymptotic limit $L\to\infty$ could be seen for
large but finite boxes. The latter should then be analyzed using the framework
of Generalized $\chi$PT \cite{gchipt,twoflav}. We have clearly in mind the
possibility to use unquenched lattice simulations, varying $N_f$ and
(finite) lattice size $L$ to eventually detect chiral phase transitions,
through the volume dependence of inverse moments (\ref{invmom}).

This article is organized as follows. In Sec.~2, we briefly review
features of Euclidean QCD and of the effective theory on a torus. Sec.~3
explains how both theories are matched to derive the original form of
Leutwyler-Smilga sum rules below $\ncrit$, before analyzing
how they are modified in the phase where the quark condensate vanishes.
In Sec.~4, we discuss the approach to the
critical point, where a competition between a small quark condensate and
higher order contributions leads to sizeable computable finite-volume effects.
Sec.~5 is devoted to the computation of the next-to-leading-order corrections to the sum
rules. We discuss in Sec.~6
how to obtain from the inverse moments an unambiguous signal indicating that
$N_f$ approaches $\ncrit$, and we discuss the interest of lattice simulations
in this framework. Sec.~7 summarizes the main results of this work.

\newpage

\section{Small mass and large volume expansion of the partition function}

\subsection{Euclidean QCD on a torus}

The Euclidean\footnote{In this paper,
all the expressions are written in the Euclidean
metric, unless explicitly stated.} QCD Lagrangian for $N_f$
light quarks reads:
\begin{equation}
{\mtca{L}}\nflav=\frac{1}{4g^2}G^a_{\mu\nu}G^a_{\mu\nu}-i\theta\nu
      -i\bar{q}D\dirac q+\bar{q}\tilde{M}{q},
\end{equation}
with the winding number density:
\begin{equation}
\nu(x)=\frac{1}{32\pi^2} G^a_{\mu\nu}(x)\tilde{G}^a_{\mu\nu}(x),
\end{equation}
and the vacuum angle $\theta$ \cite{thetavac}. The
quark mass matrix $\tilde{M}$ is of the form:
\begin{equation}
\tilde{M}=\frac{1}{2}(1-\gamma_5)M+\frac{1}{2}(1+\gamma_5)M^\dag,
\end{equation}
where $M$ is a $N_f\times N_f$ complex matrix, diagonal in a suitable quark
basis with positive real eigenvalues.

We consider the partition function of this Euclidean theory in a finite box
$L\times L \times L\times L$, large enough to neglect safely the heavy quarks:
\begin{equation}
Z^\theta(N_f)
  ={\mtca{C}}
      \int [dG] \int [d\bar\psi][d\psi] \exp\left(-\int_V d^4x\
    {\mtca{L}}\nflav\right),
\end{equation}
where $\mtca{C}$ is a normalization constant, which may depend on the
volume, but not on the mass matrix.

We impose boundary conditions on the fields, by viewing the box
as a torus and identifying $x_\mu$ and $x_\mu+n_\mu L$ (with $n_\mu$ integers):
the gluon fields have to be periodic and the quark fields
antiperiodic in the four directions, up to a gauge transformation.
The gauge fields are classified with respect to their winding number
$\nu=\int_V dx\ \nu(x)$, which is a topologically invariant integer
(related to the gauge transformation defining the periodicity of the
fields on the torus). The index theorem asserts that $\nu$ is the difference
between the number of left-handed and right-handed Dirac eigenvectors with a
vanishing eigenvalue. 

The Dirac eigenvalues satisfy a uniform bound \cite{Vafa-Witten}:
\begin{equation} \label{VW}
|\lambda _{n}[G]| < C \frac{n^{1/d}}{L} \equiv \omega_{n}.
\end{equation}
This bound means essentially that an external gauge field lowers the eigenvalues
of the free field theory \cite{paramag}. It
involves a coefficient $C$, depending only on the geometry of the
space-time manifold, but neither on $G$, $n$ or $V=L^d$.
The partition function $Z$ can be decomposed in Fourier modes over the 
winding number:
\begin{equation}
Z^\theta(N_f)=\sum_{\nu=-\infty}^\infty e^{i\nu\theta} Z_\nu(N_f). \label{Fourznu}
\end{equation}
Each projection of positive winding number $\nu$ is:
\begin{eqnarray}
Z_\nu(N_f) &=&{\mtca{C}}
      \int_\nu [dG] e^{-S_{\mtrm{YM} }[G]} \det
           (-iD\dirac+\tilde{M})\\
   &=& {\mtca{C}} \int_\nu [dG] e^{-S_{\mtrm{YM}}[G]} 
        ({\mtrm{det}}_f M)^\nu {\prod_n}'
	\left[\frac{{\det}_f(\lambda_n^2+MM^\dag)}{(\omega_n^2+\mu^2)^{N_f}}
	  \right],
    \label{diracznu}
\end{eqnarray}
$\int_\nu$ denotes the integration over the set of the gluonic
configurations with a fixed winding number $\nu$, and
$S_{\mtrm{YM}}$ is the pure gluonic action.
$\det_f M$ is the determinant
of the $N_f\times N_f$ quark mass matrix (it is replaced by
$(\det_f M^\dag)^{-\nu}$ for $\nu\leq 0$).
The primed product includes only the
strictly positive eigenvalues: its denominator
involves the Vafa-Witten bound $\omega_n$ of Eq.~(\ref{VW}) and a 
reference mass scale $\mu$ larger than any light quark mass. It represents a
convenient normalization of the determinant, such that each factor of the
primed product is lower than 1 when the quark masses tend to zero. This
normalization does not affect any observable.
We check that the quark mass matrix and the vacuum angle arise in the partition
function through the product $M\exp(i\theta/N_f)$, consistently with
the anomalous Ward identity for the singlet axial-vector current.

The partition function for a fixed positive winding number is:
\begin{eqnarray}
Z_\nu(N_f) &=& {\mtca{C}} \int_\nu [dG] 
    e^{-S_{\mtrm{YM}}[G]} 
        (\det_f M)^\nu\\
&&  \qquad \times \left({\prod_n}'
	\frac{\lambda_n^2}{\omega_n^2+\mu^2}\right)^{N_f}
  \exp\left[\left\langle{\sum_n}'
           \log\left(1+\frac{MM^\dag}{\lambda_n^2}
	             \right)\right\rangle\right],\nonumber
\end{eqnarray}
where $\langle\rangle$ denotes the trace over flavours. Provided that
the partition function is regularized, we can expand the logarithm 
for small masses (compared to the size of the box):
\begin{eqnarray}
Z_\nu(N_f) &=& {\mtca{C}} \int_\nu [dG] 
    e^{-S_{\mtrm{YM}}[G]} 
        (\det_f M)^\nu \left({\prod_n}' 
	  \frac{\lambda_n^2}{\omega_n^2+\mu^2}\right)^{N_f}
	  \label{divergence}\\
  && \qquad \times \exp\left[\langle M^\dag M\rangle \sigma_2
	        -\frac{1}{2}\langle (M^\dag M)^2\rangle \sigma_4
		  +O(M^6) \right] \nonumber\\
&=&{\mtca{C}}'_\nu
    (\det_f M)^\nu 
   \left[1+\langle M^\dag M\rangle 
          \lcr\sigma_2\rcr\nunf
	  -\frac{1}{2}\langle (M^\dag M)^2\rangle
       \lcr \sigma_4 \rcr\nunf
         \right.\label{sumgen}\\
&& \qquad\qquad\qquad \left.+\frac{1}{2}\langle M^\dag M\rangle^2
	  \lcr (\sigma_2)^2 \rcr\nunf
	 +O(M^6)
   \right]. \nonumber
\end{eqnarray}
The inverse moments are defined for each gluonic configuration
as: $\sigma_k={\sum_n}'1/\lambda_n^k$. 
The normalization factor $\mtca{C}'_\nu$ is independent of the quark mass
matrix:
\begin{equation}
{\mtca{C}}'_\nu
  ={\mtca{C}}\int_\nu [dG]e^{-S_{\mtrm{YM}}[G]}
      \left({\prod}'_n \frac{\lambda^2_n}{\omega_n^2+\mu^2}\right)^{N_f}.
\end{equation}
The average over gluonic configurations with a given winding number is
defined by
\begin{equation}
\lcr W \rcr\nunf
    =\frac{\int_\nu [dG]e^{-S_{\mtrm{YM}}[G]}
       \left({\prod}'_n \lambda^2_n\right)^{N_f}\ W}
        {\int_\nu [dG]e^{-S_{\mtrm{YM}}[G]}\left({\prod}'_n 
	  \lambda^2_n\right)^{N_f}}.
  \label{moynu}
\end{equation}
where the denominator is a normalization factor,
$\left\langle\!\left\langle 1 \right\rangle\!\right\rangle_\nu = 1$. In
Eq.~(\ref{sumgen}), this average is applied to inverse moments $\sigma_k$
that are
particularly sensitive to the infrared tail of the Dirac spectrum.
On the other hand, $\lcr\rcr\nunf$ includes a product over eigenvalues,
which should suppress the statistical weight of the lowest eigenvalues when
the number $N_f$
of massless flavours increases. The averaged inverse moments in the
exponential of Eq.~(\ref{sumgen}) could therefore exhibit a strong dependence
on $N_f$. 

(\ref{divergence}) contains several sources of divergences. Let us first
consider the gluonic configuration as a fixed external field. In the fermion
sector, sums over the Dirac spectrum may diverge because of its
ultraviolet tail. For $\lambda\to\infty$, the number of eigenvalues in
$[\lambda,\lambda+\Delta\lambda]$ is given by the free theory:
\begin{equation}
\Delta n = \frac{N_C}{4\pi^2} V|\lambda|^3 \Delta\lambda.
\end{equation}
The expected ultraviolet divergences of the inverse moments have therefore
to be subtracted. We can write:
\begin{equation}
\sigma_2 = \tilde\sigma_2 + D_2\nflav,\qquad
\sigma_4 = \tilde\sigma_4 + D_4\nflav,
\end{equation}
where the divergent part is included in $D$, and $\tilde\sigma$ is finite.
For instance, we can choose an ultraviolet cutoff $\Lambda$ and define the
integer $K$ such that $\omega_K=\Lambda$. The regularized inverse moments
then read:
\begin{equation}
\tilde\sigma_k=\sum_{n=1}^K \frac{1}{(\lambda_n)^k},
\end{equation}
and the divergent parts behave (at the \lo\  of the volume) like:
\begin{equation}
D_2\nflav \sim V\Lambda^2, \qquad D_4\nflav \sim V\ln \Lambda.
\end{equation}
These short-distance contributions are the same for all winding-number sectors.
If we perform this splitting in Eq.~(\ref{divergence}),
we obtain the regularized partition function $\tilde{Z}_\nu$ 
involving the inverse moments $\tilde\sigma$, multiplied by an exponential
factor with divergent counterterms which contribute only to the vacuum energy:
\begin{equation}
Z_\nu(N_f) = \tilde{Z}_\nu(N_f) \exp\left[D_2\nflav \langle M^\dag M\rangle
	        -\frac{1}{2} D_4\nflav\langle(M^\dag M)^2\rangle\right].
		\label{extdiv}
\end{equation}

Secondly, the product over the eigenvalues in the fermion
determinant of Eq.~(\ref{moynu}) 
needs a regularization already for a fixed gluonic
configuration. Nevertheless, for observables dominated by the lowest Dirac
eigenvalues, we expect less sensitivity to the ultraviolet tail of the
determinant. If we split the product over eigenvalues into
ultraviolet and infrared parts \cite{Descotes-Girlanda-Stern,Duncan-Eichten}:
\begin{equation}
\Delta=\Delta_{\mtrm{IR}}\Delta_{\mtrm{UV}},
\qquad \Delta_{\mtrm{IR}}
   =\prod_{n=1}^K \left(\frac{\lambda^2_n}{\omega_n^2+\mu^2}\right)^{N_f},
   \label{truncatdet}
\end{equation}
we can expect the gluonic average of the inverse moments
to depend essentially on $\Delta_{\mtrm{IR}}$, with a
weak sensitivity on $\Lambda$.

Up to now, the gauge configuration was viewed as an external
field, but the integration over the gluonic fields leads to a
third series of divergences. Fortunately, their regularization
is rather disconnected from the fermion sector \cite{`t Hooft}
(for instance, the cut-off
may be chosen independently of $\Lambda$). For the purpose of this paper,
it is sufficient to stick to
a multiplicative renormalization of the mass matrix and the Dirac eigenvalues,
\begin{equation}
M \to Z_m M, \qquad \lambda_n \to Z_m \lambda_n,
\end{equation}
inducing a multiplicative renormalization for $\lcr\sigma_k\rcr\nunf$.
We shall only consider homogeneous quantities, like ratios of inverse moments
with the same degree of homogeneity in $\lambda$: the problem of
the renormalization in the gluonic sector is therefore
discarded in the rest of this article.

\subsection{Effective Lagrangian}

For large volumes, the massive states are exponentially suppressed. The
partition function is therefore dominated by the $N_f^2-1$
pseudo-Goldstone bosons resulting from the Spontaneous Breakdown of Chiral
Symmetry and described at low energies by the Chiral Perturbation Theory
($\chi$PT). The effective Lagrangian for Goldstone bosons is
written as a double expansion in powers of the momenta $p$ and of the
quark masses $m$:
\begin{equation}
\leff=\sum_{k,l} \mtca{L}_{(k,l)}, \label{dblexp}
\end{equation}
where $\mtca{L}_{(k,l)}$ gathers all terms contributing like $p^km^l$.
In Euclidean QCD, it has been shown
that, on a large torus, the low energy constants in
$\mtca{L}_{\mtrm{eff}}$
are not affected by finite-size effects \cite{Gass-Leut2}.

If $U(x)\in \mtrm{SU}(N_f)$ collects the Goldstone fields, the partition
function is:
\begin{equation}
Z^\theta(N_f)=\int [dU] \exp\left[
     -\int_V d^4x\  
       \leff\nflav
         (U,\partial U,Me^{i\theta/N_f}) \right]. \label{effznu}
\end{equation}
In this framework, the projection on a given winding number yields
\cite{Leut-Smil}:
\begin{eqnarray}
Z_\nu(N_f)&=&\int \frac{d\theta}{2\pi}
   \ e^{-i\nu\theta} \int [dU]
     \exp\left[-\int_V d^4x\ \leff\nflav
         (U,Me^{i\theta/N_f}) \right]\\
  &=& \frac{1}{2\pi}
    \int [d\tilde U] (\det \tilde U)^\nu 
       \exp\left[-\int_V d^4x\ \leff
           \nflav(\tilde U,M) \right], 
       \label{gpintznu}
\end{eqnarray}
with $\tilde U(x)=U(x) \exp(-i\theta/N_f)$. The path integral over
$\mtrm{SU}(N_f)$ for the partition function $Z^\theta$
ends up with an integral over $\mtrm{U}(N_f)$ for $Z_\nu$. 
Because of the invariance properties of the measures
$[dU]$ and $[d\tilde U]$, we have for any $V_1,V_2\in \mtrm{U}(N_f)$:
\begin{equation}
Z_\nu(N_f|V_1MV_2)=(\det V_1V_2)^\nu Z_\nu(N_f|M).
\end{equation}
The low-energy constants in $\leff$ are volume-independent and $N_f$-dependent
order parameters. In particular, a partial restoration of chiral symmetry
would make some of them vanish. Since the relative size of these order
parameters vary with $N_f$, the organization of the double expansion
(\ref{dblexp}) depends on the phase in which the theory is
considered.

{\bf 1.}
If the number of light flavours $N_f$ is fixed below $\ncrit$,
the quark condensate $\Sigma(N_f)$ is the order parameter that dominates the
description of SB$\chi$S for sufficiently small quark masses
(or sufficiently large volumes). The \lo\  of
the effective Lagrangian involves only a kinetic term and a term linear in
the quark mass matrix:
\begin{equation}
\mtca{L}_2\nflav=\frac{1}{4}F^2(N_f)
  \langle \partial_\mu U^\dag \partial_\mu U\rangle
     -\frac{1}{2}\Sigma(N_f)
       \langle U^\dag M+M^\dag U\rangle \label{lagschipt}.
\end{equation}
$F$ is the decay constant of the Goldstone bosons and
$\Sigma(N_f)$ is the quark condensate, introduced in Sec.~\ref{introduction} in
Eqs.~(\ref{defdecons}) and (\ref{defcondqrk}).
The expansion of the effective Lagrangian is organized in this case through the
standard power counting \cite{schipt}:
$\partial \sim p$, $M \sim p^2$, so that the \nlo\  is $O(p^4)$.

{\bf 2.} On the other hand, for $N_f>\ncrit$, the
quark condensate vanishes and we cannot rely on the previous description
anymore. In this case, the leading-order Lagrangian is the sum of
the kinetic term, $\mtca{L}_{(2,0)}$, and of a term quadratic in the quark
masses, $\mtca{L}_{(0,2)}$:
\begin{eqnarray}
\mtca{L}_{(2,0)}\nflav&=&\frac{F^2(N_f)}{4}
  \langle \partial_\mu U^\dag \partial_\mu U\rangle,\\
\mtca{L}_{(0,2)}\nflav&=&-\frac{1}{4}\left[
         \mtca{A}(N_f)\langle (U^\dag M)^2+(M^\dag U)^2\rangle   
        +\mtca{Z}_S(N_f)\langle U^\dag M+M^\dag U\rangle^2\right.\\
&&\qquad \qquad \left.+\mtca{Z}_P(N_f)\langle U^\dag M-M^\dag U\rangle^2
     +\mtca{H}(N_f)\langle M^\dag M\rangle\right].\nonumber
\end{eqnarray}
$\mtca{L}_{(0,2)}$ appears in the standard $O(p^4)$ Lagrangian at the \nlo,
and the low-energy constants $\mtca{Z}_S$, $\mtca{Z}_P$, $\mtca{A}$
and $\mtca{H}$ correspond respectively
to $L_6$, $L_7$, $L_8$ and $H_2$ of Ref.~\cite{schipt}. In this phase,
the counting used to perform the expansion at higher orders is modified
\cite{gchipt}: 
$\partial \sim M \sim p$.

In the generic case $N_f\ge 3$ (the case of two flavours is commented
in App.~\ref{sec2flav}), $\mtca{A}$, $\mtca{Z}_S$ and $\mtca{Z}_P$
are order parameters of SB$\chi$S. They are related to the low-energy
behaviour of two-point correlators of the scalar and
pseudoscalar densities\footnote{Notice that contrary to the convention used
in Refs.~\cite{gchipt} and \cite{twoflav}, the decay constant $F^2$ is not
factorized in $\mtca{L}_{(0,2)}$: $\mtca{A}$, $\mtca{Z}_S$ and $\mtca{Z}_P$
carry the dimension $(\mtrm{mass})^2$.}
$S_a(x)=\bar\psi(x)t_a\psi(x)$ and
$P_a(x)=\bar\psi(x)t_a i\gamma_5\psi(x)$, where
$\{t_a\}$ are flavour matrices. $\mtca{A}$  
stems from $\langle S_a S_b-P_a P_b\rangle$.
$\mtca{Z}_S$ is given by the correlator
$\langle S_0S_0\delta_{ab}-S_aS_b\rangle$,
and $\mtca{Z}_P$ by $\langle P_0P_0\delta_{ab}-P_aP_b\rangle$:
$\mtca{Z}_S$ and $\mtca{Z}_P$ violate the Zweig rule in the scalar and
pseudoscalar channels respectively.

$\mtca{H}$ is a high-energy counterterm, which is not an order parameter
and cannot be measured in low-energy processes. Other similar counterterms
arise at higher orders: they involve only the quark mass matrix $M$, but not
the Goldstone boson fields $U$. These counterterms are needed to subtract
short-distance singularities in QCD correlation functions of quark currents.
Their general structure is dictated by the chiral symmetry, and it is
reproduced by the high-energy counterterms on the level of the effective
Lagrangian.

{\bf 3.}
For $N_f$ just below the critical point $\ncrit$, we expect a small
(but non-vanishing) condensate and a large Zweig-rule 
violation in the 
scalar sector \cite{Descotes-Girlanda-Stern}. Linear and quadratic mass
terms in the effective Lagrangian may be of comparable size. To take into
account this possibility, we include both of them in the \lo\  of the
Lagrangian:
\begin{eqnarray}
\tilde{\mtca{L}}_2\nflav&=&\frac{1}{4}\left[F^2(N_f)
  \langle \partial_\mu U^\dag \partial_\mu U\rangle
     -2\Sigma(N_f)\langle U^\dag M+M^\dag U\rangle\right.\label{laggchipt}\\
&&\quad -\mtca{A}(N_f)\langle (U^\dag M)^2+(M^\dag U)^2\rangle   
        -\mtca{Z}_S(N_f)
	   \langle U^\dag M+M^\dag U\rangle^2 \nonumber\\
&&\quad \left.-\mtca{Z}_P(N_f)\langle U^\dag M-M^\dag U\rangle^2
     -\mtca{H}(N_f)\langle M^\dag M\rangle\right].\nonumber
\end{eqnarray}
This Lagrangian can be actually viewed as the lowest order of another
systematic expansion scheme, defined by the generalized chiral counting
\cite{gchipt}: $\partial \sim M \sim B \sim O(p)$. In this case,
the \nlo\  counts as $O(p^3)$.

The standard and generalized counting rules are only two different ways of
expanding the same effective Lagrangian:
\begin{equation}
\mtca{L}_{\mtrm{eff}}=\mtca{L}_2+\mtca{L}_4+\ldots 
  =\tilde{\mtca{L}}_2+\tilde{\mtca{L}}_3+\ldots 
\end{equation}
At a given order in $p$, Generalized $\chi$PT includes terms relegated by
Standard $\chi$PT to higher orders.
At the lowest order, Eq.~(\ref{laggchipt}) can be applied even if the quark
condensate dominates. On the other hand, the Standard $\chi$PT becomes
inaccurate in the vicinity of the critical point where $\Sigma\sim 0$,
whereas Generalized $\chi$PT may be more appropriate to describe 
the transition.

\section{Leading large-volume behaviour of the inverse moments}

\subsection{Matching QCD and the effective theory} \label{matching}

If we analyze perturbatively the partition function (\ref{effznu}),
the only difference from the case of an infinite volume lies in the meson 
propagator, because of the periodic boundary conditions:
\begin{equation}
G(x)= \frac{1}{V} \sum_p \frac{e^{ipx}}{M_\pi^2+p^2},
\end{equation}
where $p_\mu=2\pi n_\mu/L$, with $n_\mu$
integers. The contribution of the mode $p=0$ in this propagator blows up
when pions become massless \cite{Gass-Leut3}.
Graphs containing such zero modes will
diverge in the chiral limit, whereas the non-zero
modes are suppressed in the large-volume limit:
the fluctuations of the zero modes
are not Gaussian and cannot be treated perturbatively. To cope with them,
we split the Goldstone boson fields in two unitary matrices: $U(x)=U_0 U_1(x)$, 
where the constant factor
$U_0$ describes the zero modes and $U_1(x)$ the remaining non-zero modes. 

In a first approximation, the Gaussian fluctuations of $U_1$ can be neglected
and the path integral in $Z$ reduces to a group integral over constant
$\mtrm{SU}(N_f)$ matrices:
\begin{equation}
Z(N_f)=\mtca{D}
    \int_{\mtrm{SU}(N_f)} [dU_0] 
      \exp\left[-V \mtca{L}\nflav_{\mtrm{eff}}(U_0,M\exp(i\theta/N_f) \right].
 \label{gpintz}
\end{equation}
where $[dU_0]$ is the Haar measure over the group, and $\mtca{D}$ a
normalization constant, independent of the mass.
The projection on a topological sector (\ref{gpintznu}) becomes:
\begin{equation}
Z_\nu(N_f)=\frac{1}{2\pi}\mtca{D}
    \int_{\mtrm{U}(N_f)} [d\tilde{U}_0] (\det \tilde{U}_0)^\nu 
       \exp\left[-V \mtca{L}\nflav_{\mtrm{eff}}(\tilde{U}_0,M) \right].
     \label{gpintznu2}
\end{equation}
To simplify the notations, we replace $\tilde{U}_0$ by $U$ in the calculations
at the \lo\  of $Z_\nu$. In addition, the $N_f$-dependence
of the low-energy constants will not be explicitly denoted from now on,
unless its presence is mandatory for understanding.

We want to expand $Z_\nu$ with respect to the size of the box
and to the quark mass matrix. Actually, Eq.~(\ref{gpintz})
tells us how to organize
this from the expansion of $\mtca{L}_{\mtrm{eff}}$. At the \lo,
the partition function will depend on a simple scaling variable
$X=ML^\kappa$. Below $\ncrit$, we have $\kappa=4$ (c.f. Eq.~(\ref{lagschipt})),
whereas the phase with a vanishing condensate yields $\kappa=2$ (c.f.
Eq.~(\ref{laggchipt})). For small $X$, the expansion of $Z_\nu$ reads:
\begin{eqnarray}
Z_\nu&=&\mtca{N}_\nu
  (\det X)^\nu [1+a_\nu \langle X^\dag X\rangle \nonumber\\
&&\qquad          +b_\nu \langle X^\dag X \rangle ^2
	  +c_\nu \langle (X^\dag X)^2 \rangle + O(X^6)
	  ],
    \label{devznu}
\end{eqnarray}
where the coefficients $\mtca{N}_\nu$, $a_\nu$, $b_\nu$, $c_\nu$ do not
depend on $M$. This expansion is valid for $\nu\geq0$: for a negative $\nu$,
$(\det X^\dag)^{|\nu|}$ arises instead of $(\det X)^\nu$.
The calculations are very similar in both cases, and our future results can be
translated for any winding number by writing $|\nu|$ instead of $\nu$.

The QCD partition function was expanded as a
polynomial in the quark masses in Eq.~(\ref{sumgen}), leading to:
\begin{eqnarray}
Z_\nu&=&\mtca{C}'_\nu
    L^{-\kappa\nu N_f} (\mtrm{det}_f X)^\nu\label{sumgenX}\\
&&\qquad   \left[1+\frac{1}{L^{2\kappa}}\langle X^\dag X\rangle
          \lcr\sigma_2\rcr\nunf
	  -\frac{1}{2L^{4\kappa}}\langle (X^\dag X)^2\rangle
       \lcr \sigma_4 \rcr\nunf
         \right.\nonumber\\
&&\qquad\quad \left.+\frac{1}{2L^{4\kappa}}\langle X^\dag X\rangle^2
	  \lcr (\sigma_2)^2 \rcr\nunf
	 +O(X^6)
   \right].\nonumber
\end{eqnarray}
By identifying the same powers of $X$ in both expansions, we obtain
relations between parameters of the effective Lagrangian and 
the leading large-volume behaviour of inverse moments.

When we compare Eqs.~(\ref{devznu}) and (\ref{sumgenX}),
we have to take into account the divergences of the 
inverse moments $\sigma_k$, as stressed in Eq.~(\ref{extdiv}):
\begin{equation}
Z_\nu=\tilde{Z}_\nu \exp\left[D_2 \langle M^\dag M\rangle
	        -\frac{1}{2} D_4\langle(M^\dag M)^2\rangle\right]
     \sim Z_\nu^{\mtrm{\chi PT}}. \label{matchdiv}
\end{equation}
These counterterms, built from traces of the quark mass matrix, are also
present in the $\chi$PT expression of the partition function. Therefore,
the divergent behaviour of the inverse moments (e.g $D_2$ for $\sigma_2$) 
is tracked by
counterterms in the $\chi$PT Lagrangian (in this case, $\mtca{H}$).
Divergence-free sum rules are found by considering linear combinations where the 
related $\chi$PT counterterms cancel.

\subsection{$N_f \ll \ncrit$: Leutwyler-Smilga sum rules}
 \label{secLS}
This case has already been treated with great details in
Ref.~\cite{Leut-Smil}. We briefly review the main steps of the derivation 
of Leutwyler-Smilga sum rules for the reader's convenience.
Eq.~(\ref{gpintznu2}) yields at the leading
order:
\begin{equation}
Z_\nu(N_f)=\frac{1}{2\pi}\mtca{D}
    \int_{\mtrm{U}(N_f)} [dU] (\det\ U)^\nu  \exp\left[\frac{\Sigma V}{2}
     \langle U^\dag M+M^\dag U\rangle \right] \label{intstand}.
\end{equation}
$VM\Sigma$ is the only parameter of the group integral, and
the scaling variable is $X=ML^4$ ($\kappa=4$).  In the general case of
an arbitrary matrix $M$, a formula for the integral (\ref{intstand}) is
discussed in Ref.~\cite{itzyk-zuber}. For our present purpose, it is
however sufficient to follow the original method described in
Ref.~\cite{Leut-Smil} to expand Eq.~(\ref{intstand}) in powers of $M$. 
We obtain the expansion coefficients $a_\nu$, $b_\nu$, \ldots
through two derivative operators, applied on both expressions of $Z_\nu$:
the group integral (\ref{intstand}) and the $X$-expansion (\ref{devznu}).
The latter gives:
\begin{eqnarray}
&&\sum_a \frac{\partial}{\partial X_a}\frac{\partial}{\partial X_a^*}
 Z_\nu = \mtca{N}_\nu(\det X)^\nu \label{sysdif1a}\\
&&\qquad \qquad \qquad \times \left\{
     \frac{N_fK}{2}a_\nu+ \langle X^\dag X \rangle
       \left[ (N_fK+1)b_\nu + (N_f+K)c_\nu \right] + O(X^4)
    \right\} \nonumber,
\end{eqnarray}
and
\begin{eqnarray}
&&\sum_{abcd} \langle t_at_bt_ct_d \rangle
 \frac{\partial}{\partial X_a}\frac{\partial}{\partial X_b^*}
  \frac{\partial}{\partial X_c}\frac{\partial}{\partial X_d^*}
 Z_\nu = \mtca{N}_\nu(\det X)^\nu     \label{sysdif1b}\\
&&\qquad \qquad \qquad 
  \times \frac{N_fK}{8} \left\{
       (N_f+K)b_\nu + (N_fK+1)c_\nu +O(X^2) \right\}, \nonumber
\end{eqnarray}
where $K=N_f+\nu$, and $X_a$ are the coordinates of $X$ on
$\{t_a\}$ ($a=0\ldots N_f^2-1$), which is a complete set of Hermitian matrices
(see App.~\ref{integrunitary}).

The same derivative operators are applied on the group integral 
(\ref{intstand}):
\begin{eqnarray}
\sum_a \frac{\partial}{\partial X_a}\frac{\partial}{\partial X_a^*}
 Z_\nu &=&\frac{1}{8}N_f\Sigma^2Z_\nu, \label{sysdif2a}\\
\sum_{abcd} \langle t_at_bt_ct_d \rangle
 \frac{\partial}{\partial X_a}\frac{\partial}{\partial X_b^*}
  \frac{\partial}{\partial X_c}\frac{\partial}{\partial X_d^*}
 Z_\nu &=& \frac{1}{256}N_f\Sigma^4Z_\nu. \label{sysdif2b}
\end{eqnarray}
Once $Z_\nu$ is replaced by its $X$-expansion (\ref{devznu}) on the right
hand-side of Eqs.~(\ref{sysdif2a}) and (\ref{sysdif2b}), these equations
yield polynomials in $X$, which are identified with Eqs.~(\ref{sysdif1a}) and
(\ref{sysdif1b}) order by order in powers of $X$. We get thus $a_\nu$, and a
linear system of two equations for $b_\nu$ and $c_\nu$.

Once $a_\nu$, $b_\nu$ and $c_\nu$ computed, 
the comparison of Eqs.~(\ref{devznu})
and (\ref{sumgenX}) leads to the Leutwyler-Smilga 
sum rules:
\begin{eqnarray}
\lcr\sigma_2 \rcr\nunf&=&a_\nu=\frac{[V\Sigma(N_f)]^2}{4K},
  \label{stdls1}\\
\lcr (\sigma_2)^2 \rcr\nunf&=&2b_\nu=
   \frac{[V\Sigma(N_f)]^4}{16(K^2-1)}, \label{stdls1bis}\\
\lcr \sigma_4 \rcr\nunf&=&-2c\nu
   =\frac{[V\Sigma(N_f)]^4}{16K(K^2-1)}. \label{stdls2} 
\end{eqnarray}
Because of $K=N_f+|\nu|$, the sum rules 
(\ref{stdls1})-(\ref{stdls2}) depend explicitly on the number of flavours,
but there is another
(implicit and unknown) dependence stemming from the quark condensate
$\Sigma(N_f)$.
No divergent counterterm is explicitly present: these sum rules are
derived from the \lo\  Lagrangian in Standard $\chi$PT, and they show
only an asymptotic behaviour, valid for $V\to\infty$. For instance, 
$\sigma_2$ and $(\sigma_2)^2$ contain divergent subleading 
terms\footnote{For this reason,
the formulae (\ref{stdls1}) and (\ref{stdls1bis}) should be
applied to finite volumes with great care.}.

\subsection{$N_f>\ncrit$: the phase with a vanishing quark condensate}
  \label{vanishing}

For $N_f>\ncrit$, the integral defining $Z_\nu$ in terms of the effective
Lagrangian (\ref{gpintznu2}) involves quadratic mass 
terms at the \lo:
\begin{eqnarray}
Z_\nu(N_f)&=&\frac{1}{2\pi}\mtca{D}
    \int_{\mtrm{U}(N_f)} [dU] (\det U)^\nu \label{znuvanish} \\ 
&&\quad \times \exp\Bigg[\frac{V}{4}
     \left\{
        \mtca{A}\langle (U^\dag M)^2+(M^\dag U)^2\rangle   
        +\mtca{Z}_S\langle U^\dag M+M^\dag U\rangle^2 \right. \nonumber\\
&&\qquad \qquad \qquad \left.+\mtca{Z}_P\langle U^\dag M-M^\dag U\rangle^2
     +\mtca{H}\langle M^\dag M\rangle\right\}\nonumber    
     \Bigg].
\end{eqnarray}
The scaling variable is now $X=ML^2$ ($\kappa=2$).
The counterterm $\mtca{H}$ has
the same structure as the divergent term $D_2$ due to $\sigma_2$ in
Eq.~(\ref{matchdiv}).
To eliminate this divergence, it is natural to introduce the $\nu$-dependent
fluctuation $\bar{\sigma}_2=\sigma_2-\lcr \sigma_2 \rcr\nunf$.
The subtraction of this quadratic divergence leads to
the loss of a single sum rule, for instance
$\lcr \sigma_2 \rcr^{(N_f)}_0$. For the other topological sectors,
we can indeed write sum rules concerning
$\lcr \sigma_2 \rcr\nunf-\lcr \sigma_2 \rcr_0^{(N_f)}$, 
since the (short-distance) divergence due to
$\mtca{H}$ is insensitive to the (topological) winding number.

Because of chirality, the integral (\ref{znuvanish})
vanishes unless the same power of $U$ and $U^\dag$ arises. 
The determinant $(\det U)^\nu$ counts as
the $\nu N_f$-th power of $U$, whereas the exponential involves only the 
square of $U^\dag$. Therefore, the phase of an odd $N_f>\ncrit$
discriminates between the topological sectors: the odd-$\nu$ sectors
are suppressed in the large-volume limit compared to the even winding
numbers (this discrimination does not occur for an even number of flavours). 
As a matter of fact, the symmetry $M\to -M$ is equivalent to
$\theta \to \theta+\pi N_f$. From the Fourier decomposition (\ref{Fourznu}),
we can directly check that the odd topological sectors have a vanishing partition
function at the \lo, provided that $N_f$ is odd. Of course, higher orders of the
effective Lagrangian (for instance $\tilde{\mtca{L}}_3$) contribute to the 
odd topological sectors, giving finally rise for $Z_\nu$ to a different volume 
dependence from the even winding numbers.

In the topologically trivial sector $\nu=0$,
$(\det U)^\nu$ disappears from the group integral and the exponential in
(\ref{znuvanish}) can be directly expanded in powers of $X$ and integrated
over $\mtrm{U}(N)$.
Using App.~\ref{integrunitary}, the computation of the lowest powers in the
$X$-expansion is straightforward, leading to the sum rules:
\begin{eqnarray}
\langle\!\langle\left
    (\bar{\sigma}_2\right)^2\rangle\!\rangle_0
 &=&\frac{V^2}{16N_f^2(N_f^2-1)}\label{ext1}\\
&& \quad\times 
[4(2N_f^2+1)(\mtca{Z}_S^2+\mtca{Z}_P^2)-8\mtca{Z}_S\mtca{Z}_P
 -8N_f\mtca{A}(\mtca{Z}_S+\mtca{Z}_P)+4N_f^2
       \mtca{A}^2],\nonumber\\
\langle\!\langle \sigma_4\rangle\!\rangle_0
  &=&\frac{V^2}{16N_f(N_f^2-1)}\label{ext2}\\
&& \quad \times [12(\mtca{Z}_S^2+\mtca{Z}_P^2)-8\mtca{Z}_S\mtca{Z}_P
     -8N_f\mtca{A}(\mtca{Z}_S+\mtca{Z}_P)+4\mtca{A}^2].
     \nonumber
\end{eqnarray}
As emphasized in the previous section, these sum rules depend on the number of
massless flavours in an explicit way, but also implicitly through the
$N_f$-dependent order parameters $\mtca{A}$, $\mtca{Z}_S$
and $\mtca{Z}_P$.

These sum rules predict a different large-volume behaviour from
the Leutwyler-Smilga sum rules
(\ref{stdls1})-(\ref{stdls2}). This agrees with our general expectation
concerning the large-volume dependence of the (suitably averaged) small Dirac
eigenvalues \cite{Stern}. The eigenvalues
accumulating like $1/L^4$ contribute to SB$\chi$S and to
the quark condensate. Correspondingly, for $N_f<\ncrit$,
the asymptotic behaviour of the sum rules is:
\begin{equation}
\left\langle\!\!\left\langle 
    {\sum_n}' \frac{1}{\lambda_n^2} \right\rangle\!\!\right\rangle_0 \sim V^2,
 \qquad 
\left\langle\!\!\left\langle {\sum_n}' \frac{1}{\lambda_n^4} 
   \right\rangle\!\!\right\rangle_0 \sim V^4.
\end{equation}
On the other hand, the $1/L^2$-eigenvalues do not contribute to the quark
condensate, but may still contribute to SB$\chi$S in the phase above
$\ncrit$, through a non-vanishing value of $F^2(N_f)$. Indeed,
Eqs.~(\ref{ext1}) and (\ref{ext2}) predict in this phase an infinite-volume
limit of $V^2$ for $\langle\!\langle\left(\bar{\sigma}_2\right)^2\rangle\!\rangle_0$
and $\langle\!\langle \sigma_4\rangle\!\rangle_0$, as expected.

\section{The approach to the critical point}

\subsection{Leading large-volume behaviour}
  \label{nearcrit}
We want now to study the intermediate case, where the
linear and the quadratic mass terms in the effective Lagrangian may compete
for some range of volumes. 
To understand which results can be expected, it is instructive to
consider first $\chi$PT in an infinite volume and to imagine that we let the quark
masses vary. If the quark condensate is (even slightly) different from zero,
we can always find sufficiently small quark masses for which the linear mass
term is dominant. When the quarks become massive, the corrections due
to the quadratic mass terms may become discernible and even 
preponderant, provided that the quark condensate is not too large to hide their
effects.

In this paper, we work in a box with a fixed large volume, and $M_\pi^2$ is 
counted as $O(1/L^4)$. The variation of the quark masses is therefore translated 
into a change of the volume. For $N_f<\ncrit$, the
Leutwyler-Smilga sum rules derived in S$\chi$PT should correctly describe
the volume-dependence of the inverse moments when $L$ tends to infinity.
However, close to the critical point and for a given value of the volume,
the quark condensate need not be large enough to make $\mtca{L}_2$,
Eq.~(\ref{lagschipt}), dominate. This could lead to significant
deviations from the asymptotic limit even for large volumes.

Hence, the \lo\  of the Lagrangian is $\tilde{\mtca{L}}_2$,
Eq.~(\ref{laggchipt}), and $Z_\nu$ reads:
\begin{eqnarray}
Z_\nu&=&\frac{1}{2\pi}\mtca{D}
    \int_{\mtrm{U}(N_f)} [dU] (\det U)^\nu  \label{znugchipt}\\ 
&&\quad \times \exp\Bigg[
  \frac{V}{4}\left\{2\Sigma\langle U^\dag M+M^\dag U\rangle
        +\mtca{A}\langle (U^\dag M)^2+(M^\dag U)^2\rangle \right. \nonumber\\
&&\qquad\qquad\qquad  \left.+\mtca{Z}_S\langle U^\dag M+M^\dag U\rangle^2
     +\mtca{Z}_P\langle U^\dag M-M^\dag U\rangle^2
     +\mtca{H}\langle M^\dag M\rangle\right\}\nonumber    
     \Bigg].
\end{eqnarray}
$X=ML^2$ remains the scaling parameter for the mass, and $\Sigma L^2$ is the
expansion variable for the quark condensate. This organizes the expansion
through the power counting $\Sigma\sim M\sim 1/L^2$, similar to G$\chi$PT.
We shall therefore consider the theory for volumes and masses so that $X$
and $\Sigma L^2$ are of order 1.

In order to evaluate (\ref{znugchipt}),
it is convenient to define the group integral
$I_\nu$ for arbitrary complex numbers $(b,\bar{b},z,\bar{z},y,a,\bar{a})$:
\begin{eqnarray}
I_\nu&=&\int_{\mtrm{U}(N_f)} \ [dU] \ (\det U)^\nu \label{definu}\\
&&\qquad  \times \exp[b\langle XU^\dag\rangle+\bar{b} \langle X^\dag U\rangle
   +z\langle XU^\dag\rangle^2+\bar{z}\langle X^\dag U\rangle^2\nonumber\\
&&\qquad \qquad \qquad +y\langle XU^\dag\rangle\langle X^\dag U\rangle+
    a\langle (XU^\dag)^2\rangle+\bar{a}\langle (X^\dag U)^2\rangle],\nonumber
\end{eqnarray}
The partition function at a fixed winding number reads:
\begin{equation}
Z_\nu=\frac{1}{2\pi}\mtca{D}\exp[h^0\langle X^\dag X\rangle]
   I_\nu(b^0,\bar{b}^0,z^0,\bar{z}^0,y^0,a^0,\bar{a}^0;X), \label{znussh}
\end{equation}
where $I_\nu$ is calculated with the values:
\begin{eqnarray}
b^0=\bar{b}^0=\frac{1}{2} L^2\Sigma,&& \label{valuelag1}\\
z^0\ =\ \bar{z}^0\ =\ \frac{1}{4}(\mtca{Z}_S+\mtca{Z}_P), & \qquad &
y^0\ =\ \frac{1}{2}(\mtca{Z}_S-\mtca{Z}_P),\\
a^0\ =\ \bar{a}^0\ =\ \frac{1}{4} \mtca{A}, & \qquad &
h^0\ =\ \frac{1}{4}\mtca{H}. \label{valuelag2}
\end{eqnarray}

$I_\nu$ is a polynomial in $(b,\bar{b},z,\bar{z},y,a,\bar{a})$, and its
derivatives are not independent:
\begin{equation}
\frac{\partial^2 I_\nu}{\partial b^2}=
      \frac{\partial I_\nu}{\partial z}\qquad
\frac{\partial^2 I_\nu}{\partial \bar{b}^2}=
      \frac{\partial I_\nu}{\partial \bar{z}}\qquad
\frac{\partial^2 I_\nu}{\partial b\partial \bar{b}}=
      \frac{\partial I_\nu}{\partial y}\label{derivinu}
\end{equation}
We expand this integral in powers of $X$, with
coefficients that are independent of the quark mass matrix:
\begin{eqnarray}
I_\nu&=&(\det X)^\nu[\alpha_\nu+\beta_\nu\langle X^\dag X\rangle
                +\gamma_\nu\langle X^\dag X\rangle^2
		+\delta_\nu\langle (X^\dag X)^2\rangle  \label{devinu}\\
  &&\qquad\qquad+\epsilon_\nu\langle X^\dag X\rangle^3
		+\eta_\nu\langle (X^\dag X)^2\rangle\langle X^\dag X\rangle
                +\kappa_\nu\langle (X^\dag X)^3\rangle
		   +O(X^8)], \nonumber
\end{eqnarray}
We identify the same powers of $X$
in the expression of $Z_\nu$ in terms of averaged inverse moments
(\ref{sumgenX}) and in its expression at the \lo\ 
of the effective Lagrangian, obtained from Eqs.~(\ref{znussh}) and
(\ref{devinu}). This leads to the sum rules:
\begin{eqnarray}
\langle\!\langle \sigma_2 \rangle\!\rangle\nunf
  &=&V\left(\frac{\beta_\nu}{\alpha_\nu}+h\right) \label{debsr}\\
\langle\!\langle \left(\sigma_2\right)^2 \rangle\!\rangle\nunf
 &=&2V^2\left(\frac{\gamma_\nu}{\alpha_\nu}+h\frac{\beta_\nu}{\alpha_\nu}
      +\frac{h^2}{2}\right)\\
\langle\!\langle \sigma_4 \rangle\!\rangle\nunf
  &=&-2V^2\frac{\delta_\nu}{\alpha_\nu}\\
\langle\!\langle \left(\bar{\sigma}_2\right)^2 \rangle\!\rangle\nunf
 &=&V^2\left(
  2\frac{\gamma_\nu}{\alpha_\nu}-\left(\frac{\beta_\nu}{\alpha_\nu}\right)^2
   \right)\\
\langle\!\langle \left(\bar{\sigma}_2\right)^3 \rangle\!\rangle\nunf
  &=&V^3\left(
      6\frac{\epsilon_\nu}{\alpha_\nu}
      -6\frac{\beta_\nu}{\alpha_\nu}\frac{\gamma_\nu}{\alpha_\nu}
      +2\left(\frac{\beta_\nu}{\alpha_\nu}\right)^3
      \right)\\
\langle\!\langle \bar{\sigma}_2\sigma_4 \rangle\!\rangle\nunf
 &=&V^3\left(-2\frac{\eta_\nu}{\alpha_\nu}
        +2\frac{\beta_\nu}{\alpha_\nu}\frac{\delta_\nu}{\alpha_\nu}\right)\\
\langle\!\langle \sigma_6 \rangle\!\rangle\nunf
  &=&3V^3\kappa_\nu \label{finsr}
\end{eqnarray}
If we know $\alpha_\nu$, $\beta_\nu$,\ldots in terms of the low-energy
constants of $\tilde{\mtca{L}}_2$, Eqs.~(\ref{debsr})-(\ref{finsr}) lead to the
desired sum rules. The high-energy counterterm $h$, which reflects the
ultraviolet divergence in $\sigma_2$, has to be eliminated. This can be obtained
if we consider the fluctuation of $\sigma_2$ over a topological sector: 
$\bar\sigma_2=\sigma_2-\lcr \sigma_2 \rcr\nunf$, as defined in
Sec.~\ref{vanishing}.

For the topologically trivial sector $\nu=0$, the computation is very
simple, following the same line as for the phase $N_f>\ncrit$.
This leads to the expansion coefficients 
(for $b=\bar{b}$, $z=\bar{z}$, $a=\bar{a}$):
\begin{eqnarray}
\alpha_0 &=& 1 \qquad \beta_0 \ =\  \frac{1}{N_f}(y+b^2)\\
\gamma_0 &=& \frac{1}{N_f(N_f^2-1)}\\
&&  \times \left\{N_f[\frac{b^4}{2}+2b^2y+2b^2z+y^2+2z^2+2a^2]-2a[b^2+2z]
            \right\}\nonumber\\
\delta_0 &=& \frac{1}{N_f(N_f^2-1)}\\
&&  \times \left\{-[\frac{b^4}{2}+2b^2z+2b^2y+y^2+2z^2+2a^2]+N_f\cdot2a[b^2+2z]\right\}
  \nonumber\\
\epsilon_0 &=&\frac{1}{N_f(N_f^2-1)(N_f^2-4)}\\
&&  \times
    \left\{6(N_f^2-2)\left[\frac{b^6}{36}+b^4\left(\frac{z}{3}+\frac{y}{4}\right)
 +b^2\left(z^2+\frac{y^2}{2}+yz\right)+\left(\frac{y^3}{6}+yz^2\right)\right]
      \right.\nonumber\\
&&\qquad\left.  +2(N_f^2+2)\left[b^2+y\right]a^2
 -12N_f\left[\frac{b^4}{6}+b^2\left(\frac{y}{2}+z\right)+yz\right]a\right\}\nonumber\\
\eta_0 &=&\frac{1}{N_f(N_f^2-1)(N_f^2-4)}\\
&&\times  \left\{-18N_f\left[\frac{b^6}{36}+b^4\left(\frac{z}{3}+\frac{y}{4}\right)
  +b^2\left(z^2+\frac{y^2}{2}+yz\right)+\left(\frac{y^3}{6}+yz^2\right)\right]
   \right.\nonumber\\
&&\qquad\left.-18N_f\left[b^2+y\right]a^2
  +12(N_f^2+2)\left[\frac{b^4}{6}+b^2\left(\frac{y}{2}+z\right)+yz\right]a\right\}
    \nonumber\\
\kappa_0 &=&
\frac{1}{N_f(N_f^2-1)(N_f^2-4)}\\
&&  \left\{24\left[\frac{b^6}{36}+b^4\left(\frac{z}{3}+\frac{y}{4}\right)
  +b^2\left(z^2+\frac{y^2}{2}+yz\right)+\left(\frac{y^3}{6}+yz^2\right)\right]
      \right.\nonumber\\
&&\qquad\left.  +4(N_f^2+2)\left[b^2+y\right]a^2
 -24N_f\left[\frac{b^4}{6}+b^2\left(\frac{y}{2}+z\right)+yz\right]a\right\}\nonumber\\
\end{eqnarray}
Before focusing on the resulting sum rules for the topologically trivial sector
$\nu=0$, we sketch the general derivation of the expansion coefficients
for an arbitrary winding number.

\subsection{Topologically non-trivial sectors: $\nu\neq 0$}
 \label{sectechnical}

Let us begin with the leading coefficient $\alpha_\nu$. Independent of $X$,
it can be computed for $X=x\cdot 1$, where $x$ is a complex
number. $\alpha_\nu$ is then given by the \lo\  of $I_\nu$ in $x$
(without any power of $x^*$), and it depends only on $(b,z,a)$. As a matter
of fact, $\alpha_\nu(b,z,a)$ can be deduced from $\alpha_\nu(b,0,a)$ 
because of the relations between the derivatives (\ref{derivinu}). 
The problem reduces to obtaining the \lo\  in $x$ of the group integral:
\begin{equation}
I^\alpha_\nu=I_\nu(b,a;x \cdot 1)
  =\int_{\mtrm{U}(N_f)} \ [dU] (\det U)^\nu 
       \exp[bx\langle U^\dag\rangle+ax^2\langle {U^\dag}^2\rangle].
\end{equation}
The Appendix~\ref{appcoefa} describes how $\alpha_\nu(b,0,a)$ is extracted
from this integral, leading to the polynomial:
\begin{equation}
\alpha_\nu(b,z=0,a)=\sum_{m=0\ldots \nu N_f/2} b^{\nu N_f-2m} a^m c_m,
\end{equation}
where $\{c_m\}$ are purely combinatorial coefficients.
Using $\partial^2 \alpha_\nu/\partial b^2=\partial \alpha_\nu/\partial z$,
we obtain the general expression of $\alpha_\nu$:
\begin{equation}
\alpha_\nu(b,z,a)=
   \sum_{l+2m+2p=\nu N_f} b^l a^m z^p \frac{(l+2p)!}{l!p!} c_m.
\end{equation}
In the limit case of a vanishing quark condensate ($b=0$), we check that
$\alpha_\nu$ (and therefore $I_\nu$)
vanishes if $\nu N_f$ is odd, in agreement with the parity
discrimination discussed in Sec.~\ref{vanishing}.

We obtain the next coefficients by applying the derivative operators
of Eqs.~(\ref{sysdif1a}) and (\ref{sysdif1b}) on both
representations of $I_\nu$: the group integral (\ref{definu}) and the
$X$-expansion (\ref{devinu}). We already know the result of the latter from
the phase $N_f\ll\ncrit$, studied in Sec~\ref{secLS}:
\begin{eqnarray}
&&\sum_a \frac{\partial}{\partial X_a}\frac{\partial}{\partial X_a^*}
 I_\nu \label{sysdif3a}\\
&&\qquad =(\det X)^\nu \left\{
     \frac{N_fK}{2}\beta_\nu+ \langle X^\dag X \rangle
       \left[ (N_fK+1)\gamma_\nu + (N_f+K)\delta_\nu \right] + O(X^4)
    \right\}, \nonumber\\
&&\sum_{abcd} \langle t_at_bt_ct_d \rangle
 \frac{\partial}{\partial X_a}\frac{\partial}{\partial X_b^*}
  \frac{\partial}{\partial X_c}\frac{\partial}{\partial X_d^*}
 I_\nu \label{sysdif3b}\\
&&\qquad \qquad
  =(\det X)^\nu \frac{N_fK}{8} \left\{
       (N_f+K)\gamma_\nu + (N_fK+1)\delta_\nu +O(X^2) \right\}. \nonumber
\end{eqnarray}

The two-derivative operator, applied on the group
integral (\ref{definu}) that defines $I_\nu$, leads to:
\begin{eqnarray}
\sum_a \frac{\partial}{\partial X_a}\frac{\partial}{\partial X_a^*} I_\nu
 &=&\left[\frac{N_f}{2}(y+b\bar b)+2a\bar a \langle X^\dag X \rangle \right.
     \nonumber\\
&& + (N_fz\bar b+a\bar b+\frac{N_f}{2}by)\frac{\partial}{\partial b}
   + (N_f\bar z b+\bar a b+\frac{N_f}{2}\bar by)\frac{\partial}{\partial \bar b}
    \nonumber \\
&& + (N_fz+a)y \frac{\partial}{\partial z}
   + (N_f\bar z+\bar a)y \frac{\partial}{\partial \bar z} \nonumber \\
&&   \left. + (2N_fz\bar z+\frac{N_f}{2} y^2+2\bar a z+2 a\bar z)
              \frac{\partial}{\partial y}\right]I_\nu. \label{twoder}
\end{eqnarray}
We can now replace $I_\nu$ by its $X$-expansion (\ref{devinu}), and
identify the resulting polynomial in $X$ with the right hand-side of
Eq.~(\ref{sysdif3a}). When we identify the coefficients of $X^0$,
we obtain $\beta_\nu$ in terms of $\alpha_\nu$ and its derivatives:
\begin{equation}
\alpha'_\nu=\frac{\partial \alpha_\nu}{\partial b},\qquad
\dot{\alpha}_\nu=\frac{\partial \alpha_\nu}{\partial a},\qquad
\alpha''_\nu=\frac{\partial^2 \alpha_\nu}{\partial b^2}
       =\frac{\partial \alpha_\nu}{\partial z}. \label{anuder}
\end{equation}
The coefficients of $\langle X^\dag X\rangle$ lead to an equality between
a linear combination of $\gamma_\nu$ and $\delta_\nu$, and
some derivatives of $\alpha_\nu$ and $\beta_\nu$ (these derivatives can
actually be rewritten only in terms of derivatives of $\alpha_\nu$, since we know
how $\beta_\nu$ is related to $\alpha_\nu$).

We follow the same line with the four-derivative operator.
Actually, when we apply the operator to the group integral (\ref{definu}),
we only need the lowest power of $X$, to compare it with (\ref{sysdif3b}).
Factors of higher degrees, similar
to $a\bar{a}\langle X^\dag X\rangle$ in Eq.~(\ref{twoder}) can be ignored,
and we obtain:
\begin{eqnarray}
&&\sum_{abcd} \langle t_at_bt_ct_d \rangle
 \frac{\partial}{\partial X_a}\frac{\partial}{\partial X_b^*}
  \frac{\partial}{\partial X_c}\frac{\partial}{\partial X_d^*}
 I_\nu  \label{fourder}\\
&&\quad =\left\{ \frac{N_f}{8}
  \left[\frac{1}{2}b^2\bar{b}^2+b^2(N_f\bar{a}+\bar{z})+\bar{b}^2(N_fa+z)
    +2b\bar{b}y\right. \right.\nonumber\\
&&\qquad\qquad\qquad 
    +y^2+2z\bar{z}+2a\bar{a}+2N_f(a\bar{z}+\bar{a}z)\biggr]\nonumber\\
&&\qquad + \frac{1}{8}\left[
    b\bar{b}(N_fby+2N_f\bar{b}z+2\bar{b}a)+\bar{b}(6N_fyz+4ay+2N_f^2ay)
       \right.\nonumber\\
&&\qquad\qquad \left.+b(2N_fy^2+4N_f^2\bar{a}z
       +4N_fz\bar{z}+4N_fa\bar{a}+4a\bar{z})\right] 
            \frac{\partial}{\partial b}\nonumber\\
&&\qquad+\frac{1}{8}\left[
	\frac{1}{2}b^2y^2+\bar{b}^2(2N_fz^2+4az)+b\bar{b}(4N_fyz+4ay)\right.
	   \nonumber\\
&&\qquad \qquad +y^2(5N_fz+(N_f^2+4)a)+4N_f^2\bar{a}z^2\nonumber\\
&&\qquad \qquad +8N_fa\bar{a}z+4N_fz^2\bar{z}+4a^2\bar{a}
	 \biggr] \frac{\partial^2}{\partial b^2}\nonumber\\
&&\qquad +\frac{1}{4}\left[2N_f\bar{b}yz^2+N_fby^2z+aby^2+4a\bar{b}yz\right]
     \frac{\partial^3}{\partial b^3}+
     \frac{1}{4}[N_fz+2a]zy^2 \frac{\partial^4}{\partial b^4}\nonumber\\
&&\qquad \left.
  +\frac{1}{4}[\bar{b}^2+2\bar{z}]a^2 \frac{\partial}{\partial a}
  +\frac{1}{2} a^2\bar{b}y \frac{\partial^2}{\partial a\partial b}
  +\frac{1}{4} a^2y^2 \frac{\partial^3}{\partial a\partial b^2}
  +O(X^2) \right\} I_\nu. \nonumber
\end{eqnarray}
We replace $I_\nu$ by its $X$-expansion on the right-hand side of this
equation. We keep only the coefficient for $X^0$ and we compare it with
Eq.~(\ref{sysdif3b}), to end up with a second
equality relating a linear combination of $\gamma_\nu$ and $\delta_\nu$
to the derivatives of $\alpha_\nu$. The resulting expressions
are listed in App.~\ref{appcoefb}, but it seems difficult to handle them in
general.

\subsection{Topologically trivial sector: $\nu=0$} \label{firstorder}

From the expansion coefficients $\alpha_0$, $\beta_0$\ldots of
Sec.~\ref{nearcrit}, we get the sum rules for the inverse moments of degree
4 and 6. If we denote $\zeta=V\Sigma^2/\mtca{A}$, 
$\bar{S}=\mtca{Z}_S/\mtca{A}$ and $\bar{P}=\mtca{Z}_P/\mtca{A}$,
the sum rules read:
\begin{eqnarray}
\langle\!\langle\left
    (\bar{\sigma}_2\right)^2\rangle\!\rangle_0\nflav
 &=&\frac{V^2\mtca{A}^2}{16N_f^2(N_f^2-1)}\label{gsr22}\\
&& \times \{\zeta^2+\zeta[4(2N_f^2+1)\bar{S}-4\bar{P}-4N_f]\nonumber\\
&& \quad +[4(2N_f^2+1)(\bar{S}^2+\bar{P}^2)-8\bar{S}\bar{P}
       -8N_f(\bar{S}+\bar{P})+4N_f^2]\}\label{sig1}\nonumber\\
\langle\!\langle \sigma_4\rangle\!\rangle_0\nflav
  &=&\frac{V^2\mtca{A}^2}{16N_f(N_f^2-1)}\label{gsr4}\\
&&\times \{\zeta^2+\zeta[12\bar{S}-4\bar{P}-4N_f]\nonumber\\
&&\quad  +[12\bar{S}^2+12\bar{P}^2-8\bar{S}\bar{P}+4
             -8N_f\bar{S}-8N_f\bar{P}]\}\nonumber\\
\langle\!\langle \sigma_6 \rangle\!\rangle_0\nflav
  &=& \frac{V^3\mtca{A}^3}{32N_f(N_f^2-1)(N_f^2-4)}\label{debl6}\\
&& \times\left\{
  \zeta^3+\zeta^2[30\bar{S}-6\bar{P}-6N_f]\right.\nonumber\\
&&\quad+\zeta[180 \bar{S}^2+36\bar{P}^2-72\bar{S}\bar{P}
          -72N_f\bar{S}+6(N_f^2+2)]\nonumber\\
&&\quad+[120\bar{S}^3-120\bar{P}^3+72\bar{S}\bar{P}^2
            -72\bar{S}^2\bar{P}-72N_f\bar{S}^2\nonumber\\
&&\quad\quad \left. +72N_f\bar{P}^2+12(N_f^2+2)\bar{S}-12(N_f^2+2)\bar{P}]
 \right\}\nonumber\\
\langle\!\langle \sigma_4\bar\sigma_2 \rangle\!\rangle_0\nflav
  &=& \frac{V^3\mtca{A}^3}{16N_f^2(N_f^2-1)(N_f^2-4)}\\
&&\times\left\{
  \zeta^3+\zeta^2[2(2N_f^2+7)\bar{S}-6\bar{P}-6N_f]\right.\nonumber\\
&&\quad +\zeta[36(N_f^2+1)\bar{S}^2+4(N_f^2+5)\bar{P}^2
             -8(N_f^2+5)\bar{S}\bar{P}\nonumber\\
&&\quad\quad -8N_f(N_f^2+5)\bar{S}+4(2N_f^2+1)]\nonumber\\
&&\quad +[24(N_f^2+1)\bar{S}^3-24(N_f^2+1)\bar{P}^3
                +8(N_f^2+5)\bar{S}\bar{P}^2\nonumber\\
&&\quad\quad-8(N_f^2+5) \bar{S}^2\bar{P}-8N_f(N_f^2+5)\bar{S}^2
                +8N_f(N_f^2+5)\bar{P}^2\nonumber\\
&&\quad\quad \left. +8(2N_f^2+1)\bar{S}-8(2N_f^2+1)\bar{P}] \right\}\nonumber\\
\langle\!\langle \left(\bar\sigma_2\right)^3 \rangle\!\rangle_0\nflav
  &=& \frac{V^3 \mtca{A}^3}{8N_f^3(N_f^2-1)(N_f^2-4)}\label{endl6}\\
&&\times\left\{
  \zeta^3+\zeta^2[6(N_f^2+1)\bar{S}-6\bar{P}-6N_f]\right.\nonumber\\
&&\quad +\zeta[6(2N_f^4-N_f^2+2)\bar{S}^2+6(N_f^2+2)\bar{P}^2\nonumber\\
&&\quad\quad-12(N_f^2+2)\bar{S}\bar{P}-12N_f(N_f^2+2)\bar{S}+9N_f^2]\nonumber\\
&&\quad +[4(2N_f^4-N_f^2+2)\bar{S}^3-4(2N_f^4-N_f^2+2)\bar{P}^3\nonumber\\
&&\quad\quad+12(N_f^2+2)\bar{S}\bar{P}^2
        -12(N_f^2+2) \bar{S}^2\bar{P}\nonumber\\
&&\quad\quad-12N_f(N_f^2+2)\bar{S}^2+12N_f(N_f^2+2)\bar{P}^2\nonumber\\
&&\quad\quad \left. +18N_f^2\bar{S}-18N_f^2\bar{P}]
 \right\}\nonumber
\end{eqnarray}
The dependence on the number of massless flavours is not limited to the
polynomials in $N_f$ explicitly present in the previous formulae, since $\Sigma$,
$\mtca{Z}_S$, $\mtca{Z}_P$ and $\mtca{A}$ are unknown functions of
$N_f$ (this dependence is here omitted for typographical convenience).
The singularities for $N_f=1$ (for $1/\lambda^4$- and $1/\lambda^6$-moments)
and $N_f=2$ (for $1/\lambda^6$-moments)
arise because some of the coefficients $\alpha_\nu$, $\beta_\nu$\ldots in
(\ref{devinu}) are not independent in these cases, and we can only write
(singularity-free) sum rules for differences between inverse 
moments of the same degree, e.g. $(\bar\sigma_2)^2-\sigma_4$ for $N_f=1$.

Notice that the scaling volume parameter $\zeta=V\Sigma^2/\mtca{A}$ and the ratios
$\bar{S}=\mtca{Z}_S/\mtca{A}$ and $\bar{P}=\mtca{Z}_P/\mtca{A}$
are invariant under the QCD renormalization group.
This invariance occurs also for
ratios of inverse moments with the same degree of homogeneity in $\lambda$:
\begin{equation}
R=\frac{\langle\!\langle\sigma_4\rangle\!\rangle_0}
    {\langle\!\langle(\bar\sigma_2)^2\rangle\!\rangle_0}, \quad
S=\frac{\langle\!\langle \left(\bar\sigma_2\right)^3 \rangle\!\rangle_0}
    {\langle\!\langle \sigma_6 \rangle\!\rangle_0}, \quad
T=\frac{\langle\!\langle \sigma_4\bar\sigma_2 \rangle\!\rangle_0}
    {\langle\!\langle \sigma_6 \rangle\!\rangle_0}, \quad
U=\frac {\langle\!\langle \sigma_4 \rangle\!\rangle_0^{3/2}}
    {\langle\!\langle \sigma_6 \rangle\!\rangle_0}.
\end{equation}

\begin{figure}[h]
\begin{center}
\includegraphics[width=11cm]{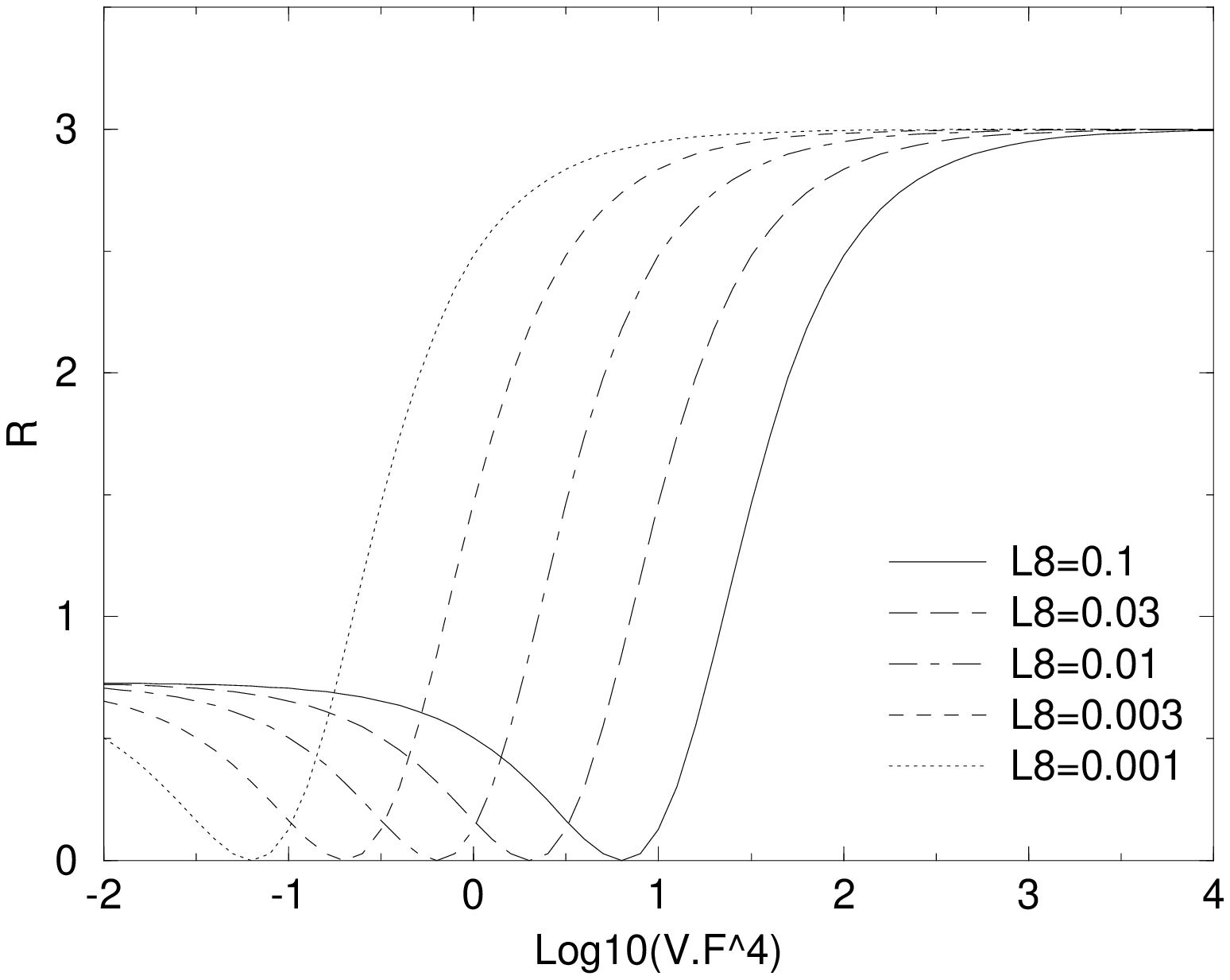}
\caption{Variations of
$R=\langle\!\langle\sigma_4\rangle\!\rangle_0/\langle\!\langle(\bar\sigma_2)^2\rangle\!\rangle_0$
as a function of the volume, measured in physical units $F_\pi^{-4}$
($N_f=3$ flavours, $\mtca{Z}_P/\mtca{A}=-1/2$ and
$\mtca{Z}_S/\mtca{A}=1/6$). The variation of $\hat{L}_8$ is only a
redefinition of the scaling parameter $\zeta$ and leads to a global shift of
the curves. The vanishing for intermediate volumes is commented in
Sec.~\ref{secposcond}.}
\label{varr1a}
\end{center}
\end{figure}

\begin{figure}[h]
\begin{center}
\includegraphics[width=11cm]{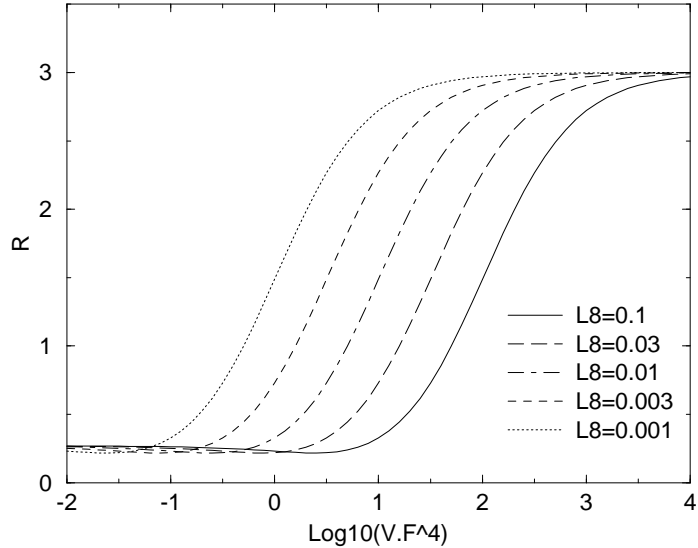}
\caption{Variations of
$R=\langle\!\langle\sigma_4\rangle\!\rangle_0/\langle\!\langle(\bar\sigma_2)^2\rangle\!\rangle_0$
as a function of the volume, measured in physical
units $F_\pi^{-4}$ ($N_f=3$ flavours, $\mtca{Z}_P/\mtca{A}=-1/2$ and
$\mtca{Z}_S/\mtca{A}=1$).}
\label{varr1b}
\end{center}
\end{figure}

We can plot (Figs.~\ref{varr1a} and \ref{varr1b})
the variation of $R$ as a function of the volume, measured in physical units
$F_\pi^{-4}$ ($F_\pi$=92.4 MeV). The scaling parameter is
$\zeta=(F_\pi^4V)/(16\hat{L}_8)$, where the dimensionless parameter $\hat{L}_8$
denotes
$(F_\pi^4\mtca{A})/(16\Sigma^2)$ (for $N_f=3$, it essentially corresponds to
the S$\chi$PT low-energy constant $L_8$ of Ref.~\cite{schipt})\footnote{In
S$\chi$PT, the constant $L_8$ depends on the renormalization scale $\mu$. At
$\mu=M_\rho$, it is estimated as $L_8^r(M_\rho)=(0.9\pm 0.3)\cdot 10^{-3}$
(see for instance Ref.~\cite{Bijnens}). 
Close to the phase transition, $L_8$ should increase
and become scale independent.}. A variation of the condensate
means a variation of $\hat{L}_8$, and consequently
a redefinition of the units used to measure the volume: this
reduces to a simple shift of the curve (to the right if $\Sigma$
decreases, to the left if it increases).

The infinite-volume limit reproduces the Leutwyler-Smilga sum rules ($R\to N_f$).
On the other hand, since the scaling volume parameter is
$\zeta=V\Sigma^2/\mtca{A}$, the limit $L\to 0$ corresponds mathematically
to a vanishing condensate for the sum rules: we recover the results of
Sec.~\ref{vanishing}. The sum rules (\ref{gsr22})-(\ref{endl6})
interpolate between these regimes.

\begin{figure}[h]      
\begin{center}
\includegraphics[width=11cm]{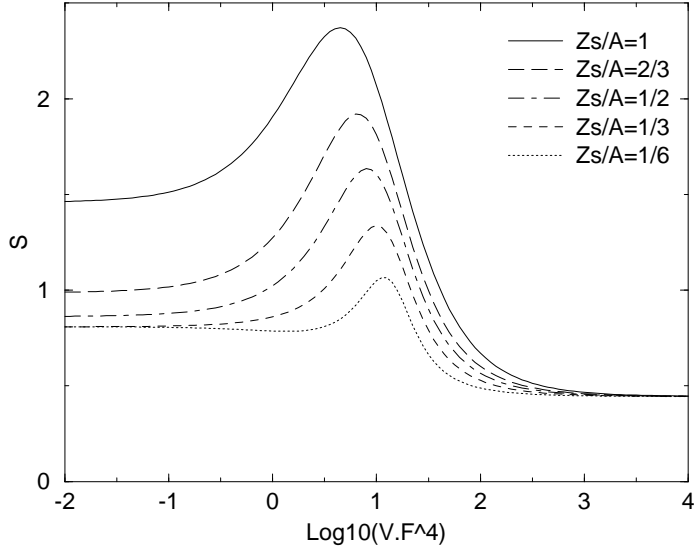}
\caption{Variations of
$S=\langle\!\langle \left(\bar\sigma_2\right)^3 \rangle\!\rangle_0/\langle\!\langle \sigma_6 \rangle\!\rangle_0$
as a function of the volume measured in physical units $F_\pi^{-4}$,
for different values of
$\bar{S}=\mtca{Z}_S/\mtca{A}$ ($N_f=3$, $\hat{L}_8=0.1$, $\bar{P}=-1/2$).
$S$ is sensitive to the parameter $\bar{S}$ even for intermediate volumes.
A different value of $\hat{L}_8$ would merely lead to a global shift of the curves.}
\label{vars1}
\end{center}
\end{figure}

The ratios $R$, $S$, $T$ and $U$
are not very sensitive to $\bar{P}$ (Zweig-rule violation in the
pseudoscalar channel) until we reach small volumes
where large corrections stemming from higher orders are expected. In the
case of $N_f=3$ flavours, the value $\bar{P}=-1/2$ is privileged, because it
guarantees the validity of the Gell-Mann--Okubo formula, independently of
the size of $\Sigma$. On the
other hand, it may be interesting to notice that some ratios are affected by
variations of $\bar{S}$ even at intermediate volumes. For instance, the
dependence of the ratio $S$ on $\bar{S}$ is plotted on Fig.~\ref{vars1}
(we choose $\hat{L}_8=0.1$, but other
values of $\hat{L}_8$ can be obtained by a simple shift of the curve).

To simplify the analysis, it may be interesting to focus on linear
combinations of the inverse moments in which the leading power of $V$ cancels.
These combinations therefore vanish in the limiting case of the Leutwyler-Smilga
sum rules (\ref{stdls1})-(\ref{stdls2}):
\begin{eqnarray}
&&N_f\lcr(\bar\sigma_2)^2\rcr\nflav_0-\lcr\sigma_4\rcr\nflav_0
 =\frac{V^2}{4N_f}
 \{2\mtca{Z}_S\Sigma^2V
    +[2\mtca{Z}_S^2+2\mtca{Z}_P^2+\mtca{A}]\}\\
&&N_f\lcr\sigma_4\bar\sigma_2\rcr\nflav_0-2\lcr\sigma_6\rcr\nflav_0
  =\frac{V^3}{8N_f(N_f^2-1)}\\
&&\quad \times\{2\mtca{Z}_S\Sigma^4V^2
    +[18\mtca{Z}_S^2+2\mtca{Z}_P^2-4\mtca{Z}_S\mtca{Z}_P
         -4N_f\mtca{Z}_S\mtca{A}+\mtca{A}^2]\Sigma^2V\nonumber\\
&&\qquad \qquad+[12(\mtca{Z}_S^3-\mtca{Z}_P^3)
      +4\mtca{Z}_S\mtca{Z}_P(\mtca{Z}_S-\mtca{Z}_P)\nonumber\\
&&\qquad \qquad \qquad \qquad -4N_f(\mtca{Z}_S^2-\mtca{Z}_P^2)\mtca{A}
      +2(\mtca{Z}_S-\mtca{Z}_P)\mtca{A}^2]\}\nonumber\\
&&N_f\lcr(\bar\sigma_2)^3\rcr\nflav_0-2\lcr\sigma_4\bar\sigma_2\rcr\nflav_0
  =\frac{V^3}{8N_f^2(N_f^2-1)}\\
&&\quad 
 \times\{2\mtca{Z}_S\Sigma^4V^2
  +[6(2N_f^2+1)\mtca{Z}_S^2+2\mtca{Z}_P^2-4\mtca{Z}_S\mtca{Z}_P
         -4N_f\mtca{Z}_S\mtca{A}+\mtca{A}^2]\Sigma^2V\nonumber\\
&&\qquad \qquad+[4(2N_f^2+1)(\mtca{Z}_S^3-\mtca{Z}_P^3)
      +4\mtca{Z}_S\mtca{Z}_P(\mtca{Z}_S-\mtca{Z}_P)\nonumber\\
&&\qquad \qquad \qquad \qquad \qquad -4N_f(\mtca{Z}_S^2-\mtca{Z}_P^2)\mtca{A}
      +2(\mtca{Z}_S-\mtca{Z}_P)\mtca{A}^2]\}\nonumber\\
&&N_f^2\lcr(\bar\sigma_2)^3\rcr\nflav_0-3N_f\lcr\sigma_4\bar\sigma_2\rcr\nflav_0
    +\lcr\sigma_6\rcr\nflav_0\\
&&\qquad \qquad =\frac{V^3}{2N_f}[3\mtca{Z}_S^2V\Sigma^2
      +2(\mtca{Z}_S^3-\mtca{Z}_P^3)]\nonumber
\end{eqnarray}
The large-volume behaviour of these combinations is particularly sensitive to
the condensate $\Sigma$ and to its fluctuation described by $\mtca{Z}_S$
(Zweig-rule violation in the scalar channel). Both these parameters are precisely
expected to be strongly affected by the vicinity of the critical point.

\subsection{Positivity conditions} \label{secposcond}

$(\bar\sigma_2)^2$, $\sigma_4$ and $\sigma_6$ are by definition positive,
and their average over any topological sector should be positive as well.
For $N_f\ll \ncrit$, this positivity is trivially satisfied by
the asymptotic behaviours predicted by the Leutwyler-Smilga sum rules
(\ref{stdls1})-(\ref{stdls2}).

When $N_f$ is near (and below) $\ncrit$,
the volume dependence of the positive inverse moments is expressed
through the sum rules of the previous section. They were derived at the
leading order, for $\nu=0$, and are functions of $\zeta$,
$\bar{S}=\mtca{Z}_S/A$ and $\bar{P}=\mtca{Z}_P/A$.
The positivity of $\lcr(\bar\sigma_2)^2\rcr\nflav_0$, $\lcr\sigma_4\rcr\nflav_0$
and $\lcr\sigma_6\rcr\nflav_0$ puts therefore constraints on the low-energy 
constants of $\tilde{\mtca{L}}_2$.

\begin{figure}[p]
\begin{center} 
\includegraphics[width=11cm]{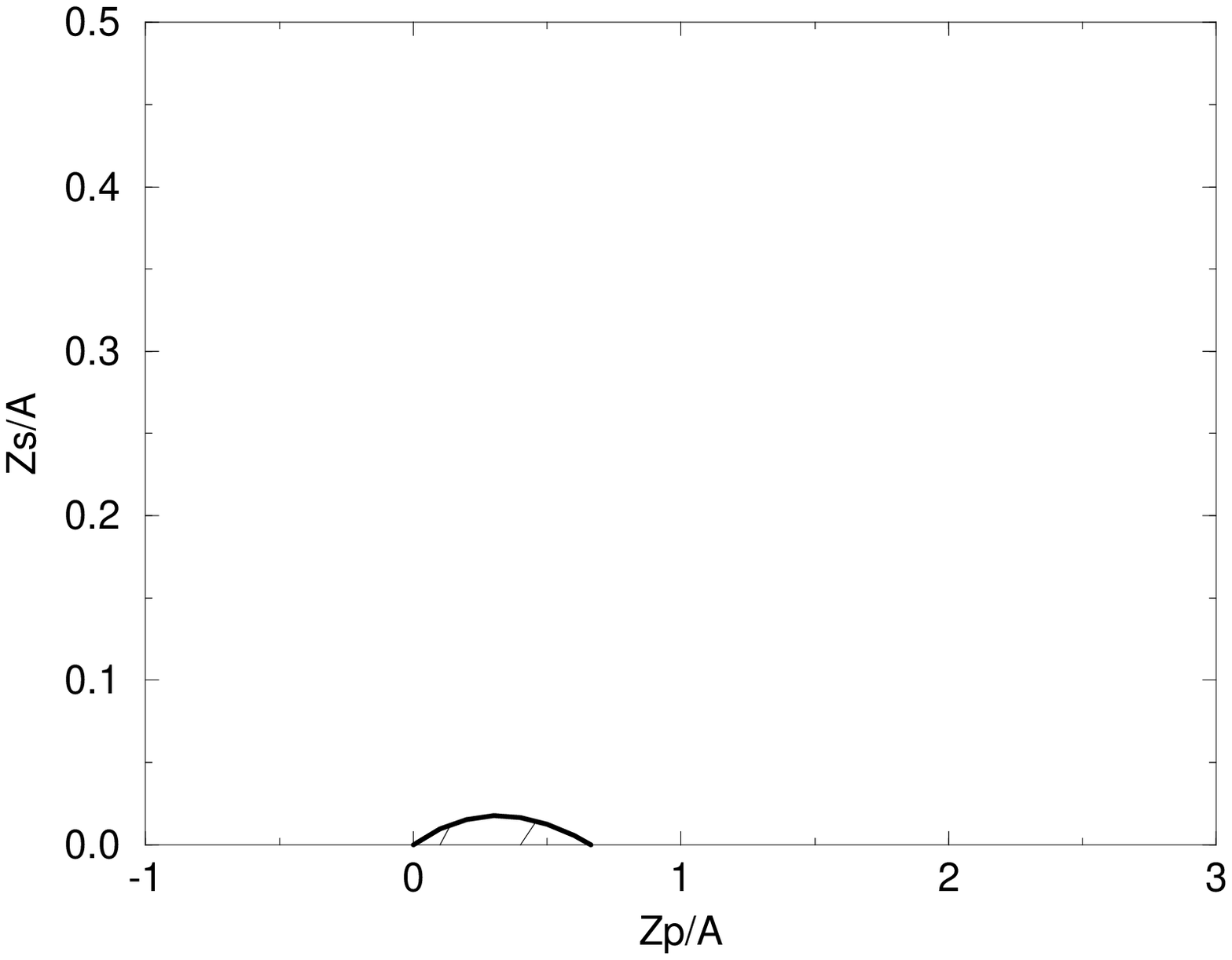}
\caption{Values of $(\mtca{Z}_S/\mtca{A},\mtca{Z}_P/\mtca{A})$
for which the sum rule (\ref{gsr22}) for $\lcr(\bar\sigma_2)^2\rcr_0^{(N_f=3)}$ 
is positive for any positive scaling parameter
$\zeta=V\Sigma^2/\mtca{A}$ (the forbidden zone is hatched).}
\label{posit1}
\end{center}
\end{figure}

\begin{figure}[h]      
\begin{center}
\includegraphics[width=11cm]{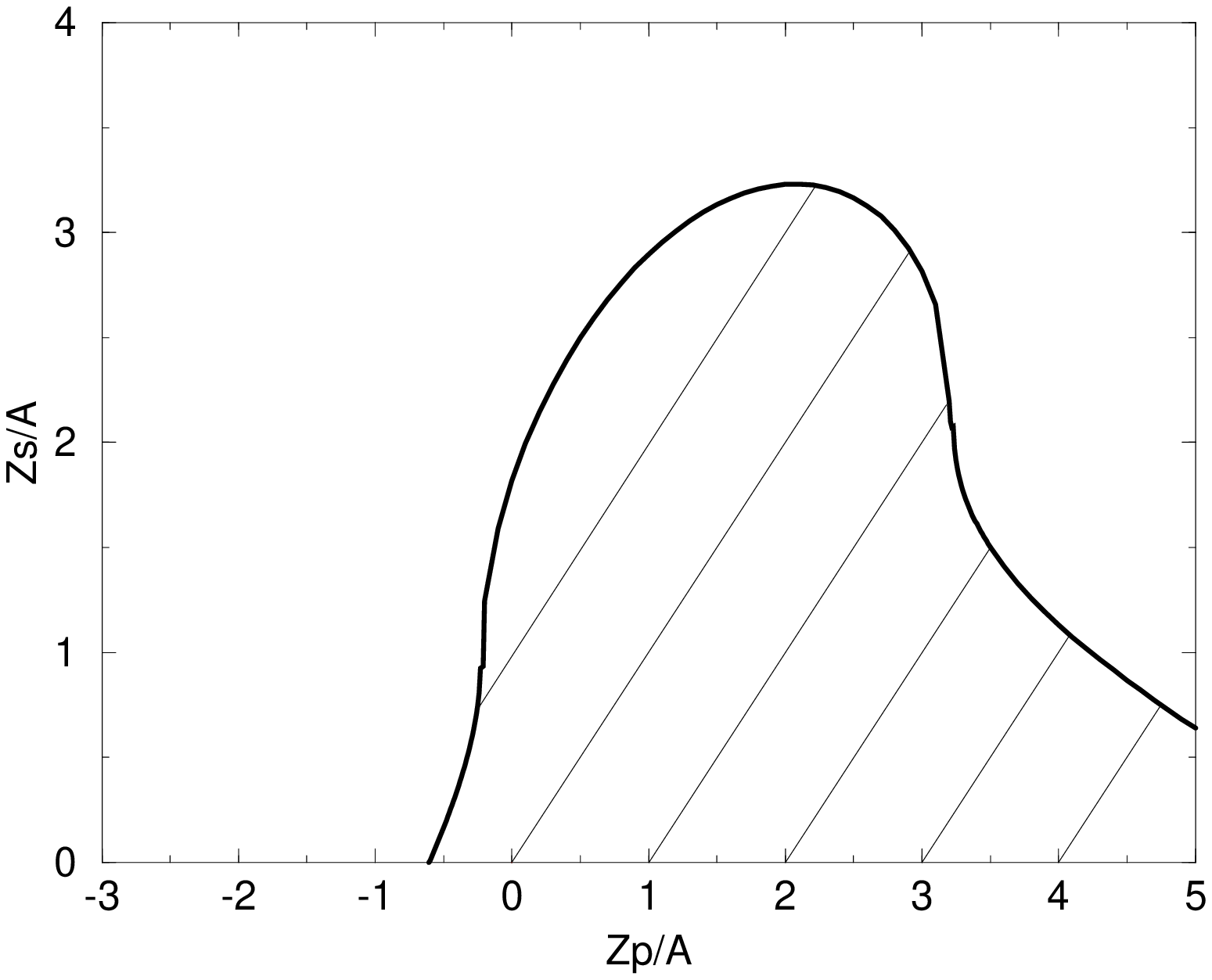}
\caption{Values of $(\mtca{Z}_S/\mtca{A},\mtca{Z}_P/\mtca{A})$,
for which the sum rule (\ref{gsr4}) 
for $\lcr\sigma_4\rcr_0^{(N_f=3)}$ is positive for
any positive scaling parameter $\zeta=V\Sigma^2/\mtca{A}$ 
(the forbidden zone is hatched).}
\label{posit2}
\end{center}
\end{figure}

\begin{figure}[h]
\begin{center}
\includegraphics[width=11cm]{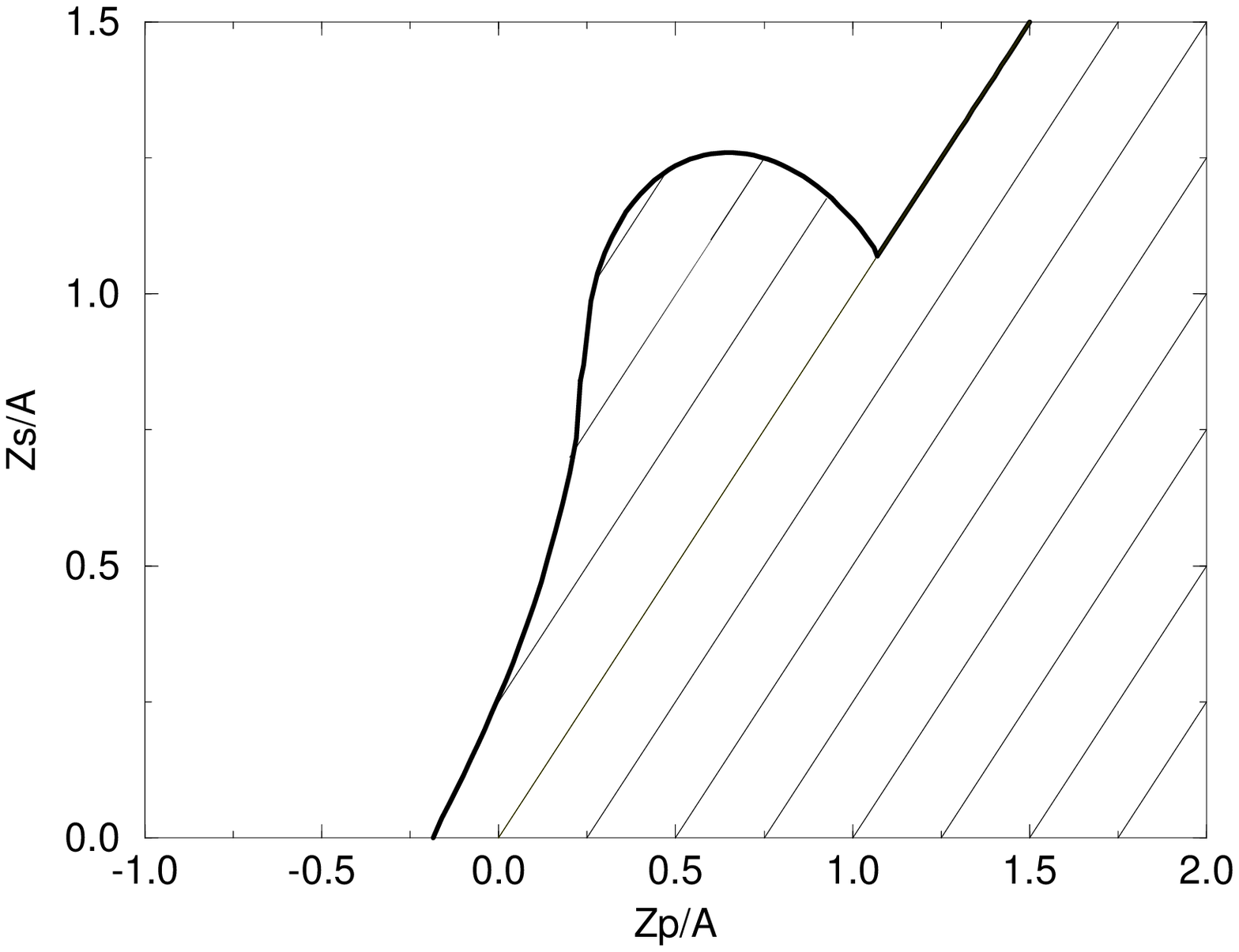}
\caption{Values of $(\mtca{Z}_S/\mtca{A},\mtca{Z}_P/\mtca{A})$,
for which the sum rule (\ref{debl6}) for 
$\lcr\sigma_6\rcr_0^{(N_f=3)}$ is positive for any positive scaling parameter
$\zeta=V\Sigma^2/\mtca{A}$ (the forbidden zone is hatched).}
\label{posit3}
\end{center}
\end{figure}

\begin{figure}[h]
\begin{center}
\includegraphics[width=11cm]{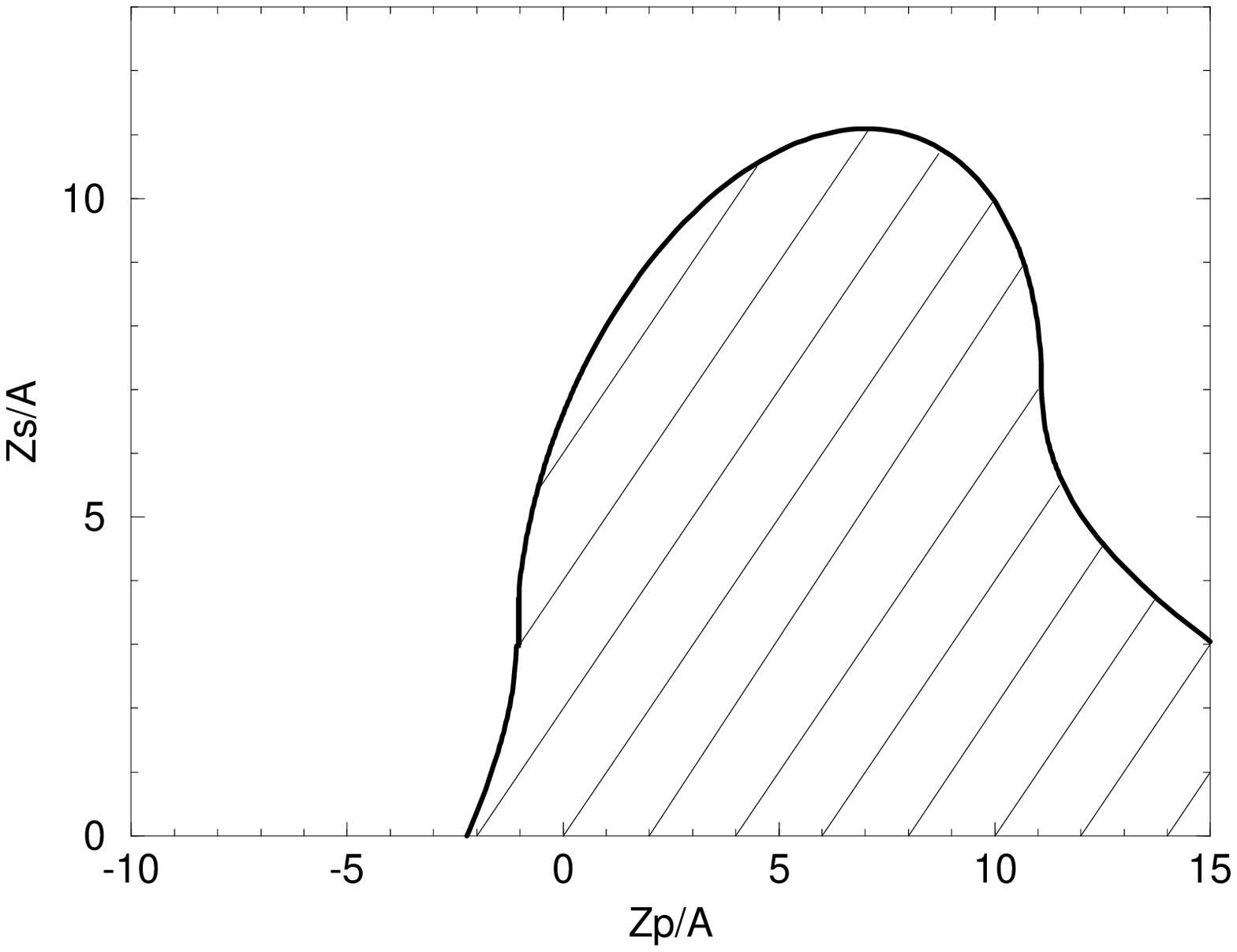}
\caption{Values of $(\mtca{Z}_S/\mtca{A},\mtca{Z}_P/\mtca{A})$
for which the sum rule (\ref{gsr4}) 
for $\lcr\sigma_4\rcr_0^{(N_f=10)}$ is positive for
any positive scaling parameter $\zeta=V\Sigma^2/\mtca{A}$ 
(the forbidden zone is hatched).}
\label{posit4}
\end{center}
\end{figure}

In the plane $(\bar{S},\bar{P})$, it is instructive to draw the domain
where each of these sum rules is positive for any value of
$\zeta=V\Sigma^2/\mtca{A}$: we demand the positivity
of a polynomial of second or third degree in $\zeta$, whose coefficients
are functions of $\bar{S}$ and $\bar{P}$ (and $N_f$). For a given number of
flavours, this procedure excludes some values of
$(\bar{S},\bar{P})$, which constitute the hatched areas
on Figs.~\ref{posit1}-\ref{posit4}. If
$\lcr(\bar\sigma_2)^2\rcr\nflav_0$ 
does not constrain $\bar{S}$ and $\bar{P}$ very much (Fig.~\ref{posit1}), 
the positivity of $\lcr\sigma_4\rcr\nflav_0$ (Fig.~\ref{posit2}) and 
$\lcr\sigma_6\rcr\nflav_0$ (Fig.~\ref{posit3}) leads to stronger conditions.
If $N_f$ increases, the excluded domains broaden, as shown on
Fig.~\ref{posit4}, compared to Fig.~\ref{posit2}.
If we suppose that $N_f=3$ is already near the critical point $\ncrit$,
and if we fix $\bar{P}=-1/2$ from the Gell-Mann--Okubo formula,
the positivity of $\lcr\sigma_4\rcr\nflav_0$ yields the condition $\bar{S} \geq 1/6$,
explaining the zero in the plot of $R$ on Fig.~\ref{varr1a}, where the
parameters $\bar{S}$ and $\bar{P}$ have been chosen on borderline of the
positivity domain of $\lcr\sigma_4\rcr\nflav_0$. 

Obviously, these areas are obtained through the leading-order approximation
to the sum rules:
the border of these domains is altered by subleading corrections, which
should become large for small volumes.
Furthermore, the pseudo-Goldstone bosons do not dominate
the partition function if the box becomes smaller than $1/\Lambda_{QCD}$.
To sum up, when we want the leading order of the sum rules to be positive
for any $\zeta$, we demand too much, and the resulting area is only an
approximation of the really allowed domain in the plane ($\bar{S},\bar{P}$).

Furthermore, these positivity plots are only relevant for a number of
flavour $N_f \sim \ncrit$. Above the phase transition,
$\lcr(\bar\sigma_2)^2\rcr\nflav_0$, $\lcr\sigma_4\rcr\nflav_0$ and
$\lcr\sigma_6\rcr\nflav_0$ are still positive, but their
large-volume behaviour is related in a different way
to the low-energy constants of the effective Lagrangian, as described in
Sec.~\ref{vanishing}. The positivity conditions stemming from the asymptotic
behaviour of $\lcr(\bar\sigma_2)^2\rcr\nflav_0$ and 
$\lcr\sigma_6\rcr\nflav_0$ are satisfied for any
$\bar{S}$ and $\bar{P}$. The only non-trivial relation is due to
the sum rule (\ref{ext2}) for $\lcr\sigma_4\rcr\nflav_0$ and reads:
\begin{equation}
(\bar{S}+\bar{P}-N_f)^2+2(\bar{S}-\bar{P})^2\geq N_f^2-1. \label{condabove}
\end{equation}
To obtain this relation, we demand the infinite-volume limit of
$\lcr\sigma_4\rcr_0\nflav$ to be positive. This limit is
predicted by the sum rule (\ref{ext2}) at the leading order.
The subleading corrections to this sum rule vanish as $L\to\infty$ and they
do not affect (\ref{condabove}). On the contrary to the previous positivity
conditions obtained
near the critical point, (\ref{condabove}) is therefore exact for $N_f>\ncrit$.

\section{Subleading corrections}

This section is devoted to the next-to-leading contributions to the sum
rules. In both phases, they behave as $1/L^2$ compared to the \lo\ 
considered so far.

\subsection{$N_f\ll \ncrit$} \label{secsubls}

The Leutwyler-Smilga sum rules were obtained at the leading
order of the effective Lagrangian $\mtca{L}_2$ in the
S$\chi$PT counting, restricted to the zero modes. The subleading corrections stem
\emph{a priori} from two sides: the non-zero modes (present already in
$\mtca{L}_2$), and the zero modes (beginning at the next-to-leading
order $\mtca{L}_4$). The first subleading corrections
turn out to be of order $1/L^2$, and they come from the non-zero modes
contributions to $\mtca{L}_2$. They can be expressed as
a (volume-dependent) renormalization of the quark condensate\footnote{This
result can be compared to the analysis performed in
Ref.~\cite{Hasen-Leut} concerning the finite size-effects arising in the
effective description of a spontaneously broken $O(N)$-symmetry.}
in the sum rules (\ref{stdls1})-(\ref{stdls2}).

The second type of subleading corrections arises from the zero-mode
contribution to $\mtca{L}_4$, quadratic in the quark mass matrix.
This Lagrangian involves,
among other terms, the counterterm $\langle M^\dag M\rangle$ corresponding
to the quadratic divergence of $\sigma_2$. Since the counting rule in this
phase is $M\sim 1/V$, these quadratic terms are suppressed by a factor 
$1/L^4$ in comparison with 
the linear term in $\mtca{L}_2$. Consequently, they appear as
next-to-\nlo\  contributions and will not be discussed here.

The non-zero modes arise in the decomposition of the Goldstone boson fields in
Sec.~\ref{matching}:
\begin{equation}
U(x)=U_0 U_1(x)=U_0 \exp\left(i\sum_{a=1}^{N_f^2-1} \xi^a(x) t_a/F\right), \qquad
\xi^a(x)=\sum_{n\neq 0} \phi^a_n 
    \exp\left(i\frac{2\pi}{L}n\cdot x\right), \label{modenon-zero}
\end{equation}
where $n_\mu$ is a four-vector whose components are integers ($n\neq 0$ means
$\sum_\mu |n_\mu|\neq 0$). The unitarity of $U_1(x)$ leads to
$\phi^a_{-n}=\left(\phi^a_n\right)^{*}$.
The fluctuations of the non-zero modes are small, leading to
the counting rule $\xi \sim \phi \sim \partial \sim 1/L$.
The leading contribution for the non-zero modes is $\partial_\mu \xi \partial_\mu \xi$
and comes from the kinetic term of $\mtca{L}_2$. It is counted
with the same power as the leading term of the
zero modes (\ref{intstand}), but it can be directly integrated and becomes a simple 
contribution to the vacuum energy \cite{Gass-Leut3}. 
At the \lo, the zero modes are
actually the only relevant degrees of freedom.

At the \nlo, the corrections from the non-zero modes are due to
the terms $\partial^2 \xi^4$ and $M \xi^2$. They are only suppressed by a
factor $1/L^2$ in comparison with the \lo\  $L^4 M$.
The partition function (\ref{gpintznu}) up to the \nlo\  is finally:
\begin{eqnarray}
Z_\nu &=&  \mtca{D}'
 \int_{\mtrm{U}(N_f)} [d\tilde{U}_0] (\det \tilde{U}_0)^\nu
     e^{-\mtca{L}_{eff}(\tilde{U}_0,X)}
   \quad\int \prod_{n>0,a} d\phi^a_n\ d(\phi^a_n)^{*}\label{nloznu}\\
&& \times \exp\left\{-\frac{L^4}{4} \sum_{n\neq 0,a,b} (\phi^a_n)^{*}
       \left[\left(\frac{2\pi}{L}\right)^2 
          \frac{n^2}{2}\delta_{ab}+Q_{ab}\right]
          \phi^b_n-T_{4\phi} +O(L^{-4})\nonumber
    \right\},
\end{eqnarray}
where the condition $n>0$ means: $n_0>0$, or $(n_0=0, n_1>0)$, or
$(n_0=n_1=0, n_2>0)$, or $(n_0=n_1=n_2=0, n_3>0)$. The
$(N_f^2-1)\times (N_f^2-1)$ matrix arises:
\begin{equation}
Q_{ab}=\frac{\Sigma}{2F^2}
 \langle(t_at_b+t_bt_a)(\tilde{U}_0^\dag M+M^\dag
 \tilde{U}_0)\rangle.
\end{equation}
$T_{4\phi}$ stands for the quartic term:
\begin{equation}
T_{4\phi}=\frac{2L^2\pi^2}{3F^2} \sum_{abcd,npqr\neq 0}
           \phi^a_n \phi^b_p \phi^c_q \phi^d_r\ n\cdot (q-p) \
     \langle t_at_bt_ct_d \rangle. \label{t4phi}
\end{equation}
In the integral over the non-zero modes in Eq.~(\ref{nloznu}),
these terms are suppressed by $1/L^2$ compared to the kinetic term.

We begin with the term $T_{4\phi}$, which
involves neither the quark mass matrix nor the zero-mode matrix
$\tilde{U}_0$. We can treat it perturbatively to perform an expansion in
powers of $1/L$, leading to:
\begin{eqnarray}
&&\int \prod_{n>0,a} d\phi^a_n\ d(\phi^a_n)^{*}\\
&&\qquad \times \exp\left\{-\frac{L^4}{4} \sum_{n\neq 0,a,b} (\phi^a_n)^{*}
       \left[\left(\frac{2\pi}{L}\right)^2 
          \frac{n^2}{2}\delta_{ab}+Q_{ab}\right]
          \phi^b_n\right\}(1-T_{4\phi}+\ldots).\nonumber
\end{eqnarray}
We should now apply Wick's theorem and contract the fields $\phi$ in $T_{4\phi}$.
We would use the propagator stemming from the kinetic term and the ``mass term''
$Q_{ab}$, where the latter is suppressed by $1/L^2$ compared to the first.
But we want only the first subleading correction due to the tadpoles arising in
$T_{4\phi}$. Since this correction is already $1/L^2$-suppressed compared to
the \lo\ of the partition function, it can be calculated
with propagators restricted to their momentum part ($Q_{ab}$ would induce
$1/L^4$-corrections). 
At the \nlo, the contribution of $T_{4\phi}$
involves neither $M$, nor $\tilde{U}_0$ (which are only present
in $Q_{ab}$): it is a global $L$-dependent term which can be factorized and
eliminated by a redefinition of the normalization constant $\mtca{D}'$.

Hence, the $1/L^2$-corrections are only due to
the ``mass term'' $Q_{ab}$ of the non-zero modes.
The partition function restricted to a given topological sector becomes:
\begin{eqnarray}
Z_\nu&=&\mtca{D}''
 \int_{\mtrm{U}(N_f)} [d\tilde{U}_0] (\det \tilde{U}_0)^\nu
      \exp(-\mtca{L}_{eff}(\tilde{U}_0,X))
   \quad\int \prod_{n>0,a} d\phi^a_n d(\phi^a_n)^{*}\\
&& \quad \times\exp\left\{-\frac{L^4}{2} \sum_{n>0,a,b} (\phi^a_n)^{*}
       \left[\frac{1}{2}n^2\left(\frac{2\pi}{L}\right)^2 
                      \delta_{ab}+Q_{ab}\right]
          \phi^b_n+O(L^{-4}) \right\}\nonumber.
\end{eqnarray}
The Gaussian integral over $\{\phi^n\}$ can now be performed:
\begin{equation}
\mtca{N} \prod_{n\neq 0} 
     \exp \left[- \frac{L^2}{4\pi^2 n^2} \mtrm{Tr\ }Q \right]
 = \mtca{N}\exp \left[-
     \frac{L^2}{4\pi^2} \left(\sum_{n\neq 0} \frac{1}{n^2}\right)
     \mtrm{Tr\ }Q \right],
\end{equation}
where $\mtca{N}$ is an $M$-independent normalization factor. The trace over
$a,b=1\ldots N_f^2-1$ leads to:
\begin{equation}
\mtrm{Tr\ }Q =\frac{N_f^2-1}{2N_f} \frac{\Sigma}{F^2}
  \langle M\tilde{U}_0^\dag
     + \tilde{U}_0 M^\dag \rangle.
\end{equation}
The integration over the non-zero modes ends up with the renormalization:
\begin{equation}
\Sigma(N_f) \to \Sigma(N_f) \left( 1 + g \frac{N_f^2-1}{2N_f} \right),
\quad
g=\frac{v}{2\pi^2F^2L^2}, \quad v={\sum}' \frac{1}{n^2}. 
\end{equation}
If we include the first subleading corrections, the sum rules
(\ref{stdls1})-(\ref{stdls2}) remain therefore correct, provided that
the parameters of the effective Lagrangian are renormalized, introducing
an additional $1/L^2$-dependence related to the regularization scheme.
In the dimensional regularization introduced by Hasenfratz and Leutwyler
\cite{Hasen-Leut}, the divergent sum $v$ becomes $-4\pi^2\beta_1$,
where $\beta_1$ is a
``shape coefficient'', related to the dimension and the geometry of the
space-time. For a four-dimensional torus, $\beta_1=0.1405$ 
(see App.~\ref{appdimreg} for further comments).

In this case, the first subleading corrections to
Eqs.~(\ref{stdls1})-(\ref{stdls2}) are summed up by the renormalization:
\begin{equation}
\Sigma(N_f) \to \Sigma_c(N_f) =
   \Sigma(N_f)\left(1-\frac{N_f^2-1}{N_f}\cdot\frac{\beta_1}{F^2L^2}\right).
\end{equation}
For instance, the relative correction $(\Sigma-\Sigma_c)/\Sigma$ remains smaller
than $\alpha$ if the box size is greater than:
\begin{equation}
L_{\mtrm{min}}=\frac{1}{F} \sqrt{\frac{N_f^2-1}{N_f}\frac{\beta_1}{\alpha}},
\end{equation}
so that, for $N_f=3$ flavours, the renormalization of $\Sigma$ in the sum
rules leads to a correction smaller than ten percent
for box sizes larger than $1.9/F$ (in the case of the dimensional
regularization).

\subsection{Near the critical point}

As before, two sources of subleading corrections should
be taken into account: the non-zero modes from $\tilde{\mtca{L}}_2$,
Eq.~(\ref{lagschipt}), and the zero modes from the next-to-leading Lagrangian
$\tilde{\mtca{L}}_3$ \cite{gchipt}:
\begin{eqnarray}
\tilde{\mtca{L}}\nflav_3&=&\frac{1}{4}
   \left\{\mtca{X}(N_f)
   \langle \partial_\mu U^\dag \partial_\mu U(M^\dag U+U^\dag M)
           \rangle\right.\\
&&\quad +\tilde{\mtca{X}}(N_f)
       \langle \partial_\mu U^\dag \partial_\mu U\rangle
         \langle M^\dag U+U^\dag M \rangle\nonumber\\
&&\quad -\mtca{R}_1(N_f)\langle(M^\dag U)^3+(U^\dag M)^3\rangle\nonumber\\
&&\quad -\mtca{R}_2(N_f)\langle(M^\dag U+U^\dag M)M^\dag M\rangle\nonumber\\
&&\quad -\mtca{R}_3(N_f)\langle M^\dag U-U^\dag M\rangle
         \langle(M^\dag U)^2-(U^\dag M)^2\rangle\nonumber\\
&&\quad -\mtca{R}_4(N_f)\langle(M^\dag U)^2+(U^\dag M)^2\rangle
          \langle M^\dag U+U^\dag M\rangle\nonumber\\
&&\quad -\mtca{R}_5(N_f)\langle M^\dag M\rangle
      \langle M^\dag U+U^\dag M\rangle\nonumber\\
&&\quad   -\mtca{R}_6(N_f)\langle M^\dag U-U^\dag M\rangle^2
          \langle M^\dag U+U^\dag M\rangle\nonumber\\
&&\left.\quad -\mtca{R}_7(N_f)
       \langle M^\dag U+U^\dag M\rangle^3 \right\}. \nonumber
\end{eqnarray}

Since the counting rule is $ML^2\sim 1$, both types of corrections are
expected to contribute at the \nlo\  $O(1/L^2)$, and could
affect the previous quadratic or cubic volume-dependence of the
sum rules.

The non-zero modes are explicitly defined by (\ref{modenon-zero}).
Like in the standard counting, their leading term in the effective Lagrangian
is the kinetic term $\partial_\mu \xi \partial_\mu \xi$, which is counted as
$O(1/L^4)$. Its contribution (at the \lo) reduces to an 
overall constant, redefining the normalization of the partition function.

The next-to-leading contributions from the non-zero modes are of
the form $B^a M^b \partial^c \xi^d$, with $2a+2b+c+d=6$,
$c$ and $d$ even, and $c\leq d$. The possible terms are $BM\xi^2$,
$M^2\xi^2$ and $\partial^2\xi^4$ from $\tilde{\mtca{L}}_2$, and
$M\partial^2\xi^2$ from $\tilde{\mtca{L}}_3$. At the \nlo\ order, the
path integral becomes:
\begin{eqnarray}
Z_\nu &=&  \mtca{D}'
 \int_{\mtrm{U}(N_f)} [d\tilde{U}_0] (\det\tilde{U}_0)^\nu
   \exp(-\mtca{L}_{eff}(\tilde{U}_0,X))
   \quad\int \prod_{n>0,a} d\phi^a_n\ d(\phi^a_n)^{*}\label{subz01}\\
&& \times \exp\Bigg\{-\frac{L^4}{2} \sum_{n>0,a,b} (\phi^a_n)^{*}
       \left[n^2\left(\frac{2\pi}{L}\right)^2 
          \left(\frac{1}{2}\delta_{ab}+\tilde{P}_{ab}\right)
	     +\tilde{Q}_{ab}\right]
          \phi^b_n
-T_{4\phi} +O\left(\frac{1}{L^4}\right) \Bigg\},\nonumber
\end{eqnarray}
with the $(N_f^2-1)\times (N_f^2-1)$ matrices:
\begin{eqnarray}
\tilde{P}_{ab}
  &=&\frac{1}{2F^2}\left[\mtca{X}
      \langle\{t_a,t_b\}(\tilde{U}_0^\dag M+M^\dag \tilde{U}_0)\rangle
         +\tilde{\mtca{X}} \delta_{ab} 
	   \langle \tilde{U}_0^\dag M+M^\dag \tilde{U}_0\rangle\right],\\
\tilde{Q}_{ab}&=&\frac{1}{F^2}\Bigg[\frac{\Sigma}{2}
 \langle(t_at_b+t_bt_a)(\tilde{U}_0^\dag M+M^\dag
 \tilde{U}_0)\rangle\nonumber\\
&&\quad +\mtca{A} \langle t_a \tilde{U}_0^\dag Mt_b \tilde{U}_0^\dag M
       + t_a M^\dag \tilde{U}_0t_b M^\dag \tilde{U}_0 \rangle\nonumber\\
&&\quad +\mtca{Z}_S
          \langle t_a(\tilde{U}_0^\dag M-M^\dag \tilde{U}_0)\rangle
          \langle t_b(\tilde{U}_0^\dag M-M^\dag \tilde{U}_0)\rangle\nonumber\\
&&\quad +\mtca{Z}_P
          \langle t_a(\tilde{U}_0^\dag M+M^\dag \tilde{U}_0)\rangle
          \langle t_b(\tilde{U}_0^\dag M+M^\dag \tilde{U}_0)\rangle\Bigg].
\end{eqnarray}
The quartic term $T_{4\phi}$ remains identical to its expression in the
standard case (\ref{t4phi}) and
it stems from the kinetic term of $\tilde{\mtca{L}}_2$, whereas $\tilde{P}$
is due to $\tilde{\mtca{L}}_3$ and  $\tilde{Q}$ to
the non-derivative part of $\tilde{\mtca{L}}_2$. In Eq.~(\ref{subz01}),
the contributions of these three terms are
suppressed by $1/L^2$, compared to the kinetic term:
$\pi^2L^2\sum_{n>0,a} n^2 |\phi^a_n|^2$.

For the same reasons as in the previous section, the integration of $T_{4\phi}$
leads at this order to a term independent of $M$ and $\tilde{U}_0$, which
merely redefines the overall normalization constant $\mtca{D}$.
At the \nlo, the partition function for a given winding number reads:
\begin{eqnarray}
Z_\nu&=&\mtca{D}''
 \int_{\mtrm{U}(N_f)} [d\tilde{U}_0](\det\tilde{U}_0)^\nu
   \exp(-\mtca{L}_{eff}(\tilde{U}_0,X))
   \quad\int \prod_{n>0,a} d\phi^a_n d(\phi^a_n)^{*}\label{subz02}\\
&& \quad \times\exp\left\{-\frac{L^4}{2} \sum_{n>0,a,b} (\phi^a_n)^{*}
       \left[\frac{1}{2}n^2\left(\frac{2\pi}{L}\right)^2 
                      (\delta_{ab}+2\tilde{P}_{ab})+\tilde{Q}_{ab}\right]
          \phi^b_n \right\},\nonumber
\end{eqnarray}
which yields after the integration over $\phi$:
\begin{eqnarray}
&& \mtca{N} \prod_{n\neq 0} \exp \left[- \mtrm{Tr\ }\tilde{P} 
    - \frac{L^2}{4\pi^2 n^2} \mtrm{Tr\ }\tilde{Q} \right]\\
&& \qquad = \mtca{N}
   \exp \left[- \left(\sum_{n\neq 0} 1\right) \mtrm{Tr\ }
    \tilde{P} -
     \frac{L^2}{4\pi^2} \left(\sum_{n\neq 0} \frac{1}{n^2}\right)
     \mtrm{Tr\ }\tilde{Q} \right],\nonumber
\end{eqnarray}
where $\mtca{N}$ is an $M$- and $\tilde{U}_0$-independent
normalization factor. The traces are taken over the indices $a,b=1\ldots N_f^2-1$:
\begin{eqnarray}
\mtrm{Tr\ }\tilde{P} &=&\frac{1}{F^2L^2}
    \frac{N_f^2-1}{2N_f} 
    (\mtca{X}+N_f\tilde{\mtca{X}}) 
      \langle X\tilde{U}_0^\dag + \tilde{U}_0 X^\dag\rangle\\
\mtrm{Tr\ }\tilde{Q} &=&\frac{1}{2F^2L^4}
 \left[
 \Sigma L^2 \frac{N_f^2-1}{N_f}
     \langle X\tilde{U}_0^\dag + \tilde{U}_0 X^\dag \rangle
 +\left(\mtca{A}-\frac{\mtca{Z}_S+\mtca{Z_P}}{N_f}\right)
       \left(\langle X\tilde{U}_0^\dag\rangle^2+
           \langle \tilde{U}_0X^\dag\rangle^2 \right)
 \right.\nonumber\\
&& \qquad \qquad +\left(\mtca{Z}_S+\mtca{Z_P}-\frac{\mtca{A}}{N}\right)
       \langle (X\tilde{U}_0^\dag)^2+(\tilde{U}_0X^\dag)^2\rangle\\
&& \qquad \qquad \left. +\frac{2(\mtca{Z}_S-\mtca{Z_P})}{N_f}
                 \langle X\tilde{U}_0^\dag\rangle
          \langle X^\dag \tilde{U}_0\rangle
    -2(\mtca{Z}_S-\mtca{Z_P}) \langle X^\dag X \rangle
    \right], \nonumber
\end{eqnarray}
The integration over the non-zero modes ends up with a term of the
same structure as $\tilde{\mtca{L}}_2$, i.e. it renormalizes
the parameters of the Lagrangian in the sum rules:
\begin{eqnarray}
\Sigma(N_f) &\to& \Sigma(N_f) 
   + g \frac{N_f^2-1}{2N_f} \Sigma(N_f) + \frac{2u}{F^2V}\frac{N_f^2-1}{2N_f}
   [\mtca{X}(N_f)+N_f\tilde{\mtca{X}}(N_f)],\\
\mtca{A}(N_f) &\to& \mtca{A}(N_f) 
    + g \left[\mtca{Z}_S(N_f)+\mtca{Z}_P(N_f)-\frac{\mtca{A}(N_f)}{N_f}\right],\\
\mtca{Z}_S(N_f) 
    &\to& \mtca{Z}_S(N_f) + g \left[\frac{\mtca{A}(N_f)}{2}
       -\frac{\mtca{Z}_P(N_f)}{N_f}\right],\\
\mtca{Z}_P(N_f) 
    &\to& \mtca{Z}_P(N_f) + g \left[\frac{\mtca{A}(N_f)}{2}
       -\frac{\mtca{Z}_S(N_f)}{N_f}\right],\\
\mtca{H}(N_f) & \to & \mtca{H}(N_f) 
    - g\cdot 2[\mtca{Z}_S(N_f)-\mtca{Z}_P(N_f)],
\end{eqnarray}
with the sums to be regularized:
\begin{equation}
g=\frac{v}{2\pi^2F^2L^2}, \qquad v={\sum}' \frac{1}{n^2}, \qquad u={\sum}'1.
\end{equation}
If we consider the dimensional regularization, we get $g=-2\beta_1/F^2L^2$
and $u=0$ (see App.~\ref{appdimreg}).

With the counting $\Sigma L^2\sim ML^2 \sim 1$, the first subleading corrections
stem also from the zero modes in $\tilde{\mtca{L}}_3$: they contain therefore
the low-energy constants $\mtca{R}_i$.
If we consider the topological sector $\nu=0$, the resulting corrections are
quite simple to compute. When we expand $Z_0$  (restricted to the zero modes)
as a polynomial $X$,
the integrals with different powers of $U$ and $U^\dag$ vanish. In
particular, the terms from $\tilde{\mtca{L}}_3$ involve odd powers of the
meson matrix and have to be combined with
$\Sigma\langle X^\dag U+U^\dag X\rangle$.
The resulting corrections are therefore $\Sigma\mtca{R}_i$ and
are counted as $O(1/L^2)$.

For $\nu=0$ the final form of the sum rules, including the first subleading corrections,
is:
\begin{eqnarray}
\langle\!\langle (\bar\sigma_2)^2\rangle\!\rangle_0&=&
  \frac{V^2\mtca{A}^2}{16N_f^2(N_f^2-1)}
        \left[s_2^0+s_2^{\mtca{R}}+s_2^r
	\right],\label{sig2cor}\\
\langle\!\langle \sigma_4\rangle\!\rangle_0&=&
        \frac{V^2\mtca{A}^2}{16N_f(N_f^2-1)}
        \left[s_4^0+s_4^{\mtca{R}}+s_4^r
	\right],\label{sig4cor}
\end{eqnarray}
where $s_k^0$ is the leading term, already calculated in
Sec.~\ref{firstorder}, $s_k^{\mtca{R}}$ collects the terms from the zero modes in
$\tilde{\mtca{L}}_3$, and $s_k^r$ is due to the renormalization of
$\tilde{\mtca{L}}_2$ induced by the non-zero modes. The result is:
\begin{eqnarray}
s_2^0&=&\zeta^2+\zeta[4(2N_f^2+1)\bar{S}-4\bar{P}-4N_f]\\
&& \quad +[4(2N_f^2+1)(\bar{S}^2+\bar{P}^2)-8\bar{S}\bar{P}
            -8N_f(\bar{S}+\bar{P})+4N_f^2],\nonumber\\
s_2^{\mtca{R}}&=&\frac{\Sigma}{\mtca{A}^2}
    [16N_f(\mtca{R}_3-\mtca{R}_4)+8N_f(N_f^2-1)\mtca{R}_5
           -16N_f^2\mtca{R}_6+48N_f^2\mtca{R}_7],\\
s_2^r &=& \frac{g}{N_f} \left\{ 2(N_f^2-1)\zeta^2
     +\zeta[8N_f^2(N_f^2-1)\bar{S}-16N_f^2 \bar{P}+8N_f]\right.\nonumber\\
&& \qquad +[-8(N_f^2-1)(\bar{S}^2+\bar{P}^2)-16(3N_f^2+1)\bar{S}\bar{P}\nonumber\\
&& \qquad\quad \left. +16N_f(N_f^2+1)(\bar{S}+\bar{P})-16N_f^2]\right\}\\
&& +u\cdot 4\frac{N_f^2-1}{N_f}
     \frac{\Sigma(\mtca{X}+N_f\tilde{\mtca{X}})}{F^2\mtca{A}}
     \left\{\zeta+2\left[(2N_f^2+1)\bar{S}-\bar{P}-N_f\right] \right\},\nonumber
\end{eqnarray}
and
\begin{eqnarray}
s_4^0&=&\zeta^2+\zeta[12\bar{S}-4\bar{P}-4N_f]\\
   &&\quad  +[12\bar{S}^2+12\bar{P}^2-8\bar{S}\bar{P}
                  -8N_f\bar{S}-8N_f\bar{P}+4],\nonumber\\
s_4^{\mtca{R}}&=&\frac{\Sigma}{\mtca{A}^2}
         [-8(N_f^2-1)\mtca{R}_2+16N_f\mtca{R}_3-16N_f\mtca{R}_4
	   -16\mtca{R}_6+48\mtca{R}_7],\\
s_4^r &=&
   \frac{g}{N_f} \left\{
    2(N_f^2-1)\zeta^2
       +\zeta[8(N_f^2-1)\bar{S}-8(N_f^2+1)\bar{P}-4N_f(N_f^2-3)]\right.\nonumber\\
&& \qquad +[-8(N_f^2-1)(\bar{S}^2+\bar{P}^2)-16(N_f^2+3)\bar{S}\bar{P}\nonumber\\
&& \qquad \quad \left.+32N_f(\bar{S}+\bar{P})-8(N_f^2+1)]
    \right\}\\
&& +u \cdot 4\frac{N_f^2-1}{N_f} 
   \frac{\Sigma(\mtca{X}+N_f\tilde{\mtca{X}})}{F^2\mtca{A}}
   \left\{\zeta+2\left[3\bar{S}-\bar{P}-N_f\right] \right\}.\nonumber
\end{eqnarray}

It is worth commenting the above results: in the vicinity of the critical
point, characterized by the counting $\Sigma L^2\sim ML^2\sim 1$, all the terms
of the leading contribution $s_k^0$ are of the same order 1.
$s_k^{\mtca{R}}$ and $s_k^r$ collect all the next-to-leading contributions,
which are counted as $O(1/L^2)$. Consequently, for a fixed value of the
condensate $\Sigma$, the inverse moments $\lcr(\bar\sigma_2)^2\rcr_0\nflav/V^2$
and $\lcr\sigma_4\rcr_0\nflav/V^2$ can be expressed in the form
$\sum_{n=-1}^4 a_n L^{2n}$. The even powers $n=2,4$ are the original leading
terms, whereas the odd powers $n=-1,1,3$ arise from the next-to-leading
corrections due to the non-zero modes. Hence, this type of
correction does not mix with the leading contribution as far as the
volume dependence is concerned.

This is not true for $s_k^{\mtca{R}}$, which stems from the zero-mode
contribution of
$\tilde\mtca{L}_3$. They modify the constant term $n=0$ of the sum rules, and
may be considered as small to the extent that $\Sigma$ is small (let us
recall that the dimensional estimate of the low-energy constants
$\mtca{R}_i$ leads to $\mtca{R}_i\sim F^2/\Lambda_H$ with
$\Lambda_H\sim$~1~GeV). Of course, close to the critical point, one precisely
expects $\Sigma$ to become small.

In the case of the $\lambda^{-6}$-sum rules (\ref{debl6})-(\ref{endl6}), the
situation is similar, but now, the constants $\mtca{R}_i$ already affect
the coefficient of $\zeta^1$ in the sum rules (they also change the constant term
$\zeta^0$).

\section{Extraction of particular low-energy constants}

Near the critical point $\ncrit$, we would like to exploit
the sum rules for $\lcr(\bar\sigma_2)^2\rcr\nflav_0$ and
$\lcr\sigma_4\rcr\nflav_0$ in order to isolate
particular ratios of low-energy constants present in $\tilde{\mtca{L}}_2$.
In particular, it would be interesting to obtain a ratio with a specific
sensitivity to the phase transition.
To reach this goal, it is preferable to eliminate the
next-to-leading corrections, which involve either unknown parameters like
$\mtca{R}_i$ or regularization-dependent quantities like $g$.
As already pointed out, Eqs.~(\ref{sig2cor}) and (\ref{sig4cor}) can be
viewed as expansions in the variables $\Sigma L^2$ and $ML^2$. Hence, they
are even functions of $L$.

\subsection{Varying the size of the box}

To exploit the structure of the sum rules at the \nlo, 
it is therefore interesting to introduce the derivative-like operator:
\begin{eqnarray}
\delta_a[F](L)&=&\frac{a^2}{8L\left(L^2-\frac{a^2}{4}\right)}\label{defda}\\
&&\quad \times\left\{\left(L-\frac{a}{2}\right)F(L+a)
         +\left(L+\frac{a}{2}\right)F(L-a)
	 -2LF(L)\right\},\nonumber
\end{eqnarray}
where $a$ is an arbitrary parameter. If we consider an even monomial
$F(L)=L^{2k}$, $\delta^a[F]$ is an even polynomial of
degree $L^{2k-4}$. We obtain for the first powers:
\begin{eqnarray}
L^0 \to 0, &\qquad &  L^2 \to 0, \qquad L^4 \to a^4,\\
L^6 \to a^4(3L^2+2a^2),&\qquad& L^8 \to a^4(6L^4+12L^2a^2+3a^2).
\end{eqnarray}

If we denote $t_2=\lcr(\bar\sigma_2)^2\rcr_0\nflav/V^2$ and
$t_4=\lcr\sigma_4\rcr_0\nflav/V^2$, we have:
\begin{eqnarray}
\delta_a[N_ft_2-t_4](L)&=&\frac{a^4}{2N_f}\Sigma^2\mtca{Z}_S+O(1/L^6),
  \label{latt1}\\
\delta_a[\delta_a[t_2]](L)&=&\frac{3a^8}{8N_f^2(N_f^2-1)}\Sigma^4+O(1/L^{10}).
 \label{latt2}
\end{eqnarray}

In order to get a quantity which is invariant with respect to the QCD
renormalization group, we take the ratio of these two sum rules:
\begin{equation}
\rho=\frac{\delta_a[\delta_a[t_2]]}{\delta_a[N_ft_2-t_4]}(L)
   =\frac{3a^4}{4N_f(N_f^2-1)}\cdot 
         \frac{\Sigma^2}{\mtca{Z}_S}+O(1/L^6). \label{latt3}
\end{equation}
The evaluation of $\rho$ requires the knowledge of $t_2$ and $t_4$ for five
different box sizes: $L-2a$, $L-a$, $L$, $L+a$, $L+2a$. Notice that $a$ is
not required to be large; it is sufficient to have $L-2a$ much bigger
than $1/\Lambda_H$. On the other hand, for too small $a$, the difference
operator $\delta_a$ may be too sensitive to numerical errors.

For a discretized torus (a lattice), we can put $L=na$ with $n$ integer and
$a$ the lattice spacing. Eqs.~(\ref{latt1})-(\ref{latt3}) remain true,
and the comparison of different volumes is translated into the
evaluation of the inverse moments on lattices with the same spacing,
but with various numbers of sites. The powers in the lattice
spacing $a$ on the right-hand side of Eqs.~(\ref{latt1})-(\ref{latt3})
reflect merely the dimension of the involved quantities.

Eqs.~(\ref{latt1})-(\ref{latt3})
are independent of the \nlo\ contributions: this allows to
consider smaller volumes than previously stated
(for instance, the estimate $L>1.9/F$ of Sec.~\ref{secsubls}, based on our
\nlo\ analysis does not necessarily apply to the sum rule (\ref{latt3})). 
The volume-independence of Eqs.~(\ref{latt1})-(\ref{latt3}) could already
be seen for smaller volumes. Moreover, the 
inverse moments must fulfill another non-trivial consistency relation:
\begin{equation}
\rho'=\frac{\delta_a[\delta_a[t_2]]}{\delta_a[\delta_a[t_4]]}(L)
   =\frac{1}{N_f}+O(1/L^{10}). \label{latt4}
\end{equation}
The ratio $\rho$ is
invariant under the QCD renormalization group and
its variations with $N_f$ could reflect the
proximity of the critical point in a particularly clean way,
as discussed in the next section.

\subsection{Relevance of the ratio $\mtca{Z}_S/\Sigma^2$}

We have argued in a previous paper
\cite{Descotes-Girlanda-Stern} that the approach to
$\ncrit$ could result into a large Zweig-rule violation in the scalar channel.
Let us recall briefly the argument.
We consider the chiral limit for the first $N_f$ light flavours of common
mass $m\to 0$, and denote by $s$ the $(N_f+1)$-th quark, whose mass $m_s$ is
non-zero, but still considered as light compared to the scale of the theory
(real QCD corresponds to $N_f=2$). Here, we typically consider $N_f$ such
that $N_f+1<\ncrit\leq N_f+2$. $\Sigma(N_f)$ is a function of $m_s$,
with the derivative:
\begin{equation} \label{der}
\frac{\partial}{\partial m_s} \Sigma(N_f) = 
\lim_{m\to 0} \int\ dx \langle \bar u u(x) \bar s s(0)\rangle^{c} 
    \equiv \Pi_{Z}(m_s),
\end{equation}
where the superscript $c$ stands for the
connected part. Since $\Sigma(N_f) \to \Sigma(N_{f}+1)$
for $m_s \to 0$, one can write:
\begin{equation}
\Sigma(N_f) = \Sigma(N_{f}+1) + \int_{0}^{m_s}\!\!d\mu\ \Pi_{Z}(\mu)
  =\Sigma(N_{f}+1) + m_s Z_{\mtrm{eff}}(m_s) + O(m_s^2\log m_s),
  \label{diff}
\end{equation}
where $Z_{\mtrm{eff}}(m_s)$, defined in Ref.~\cite{Descotes-Girlanda-Stern},
is essentially $2\mtca{Z}_S(N_f+1)$,
up to corrections of the order $(\Sigma(N_f+1))^2$,
which are small in the vicinity of the critical point.
Close to $\ncrit$, the condensate term
need not dominate the expansion (\ref{diff}) in powers of $m_s$, 
due to the suppression of
$\Sigma(N_f+1)$. The large variation of the quark condensate from $N_f$ to
$N_f+1$ is then reflected by a large value of $\mtca{Z}_S(N_f+1)$, 
related to the Zweig-rule violation in the $0^{++}$ channel.
Once expressed through the Dirac spectrum,
$\Sigma$ can be interpreted as the average density of small eigenvalues,
whereas $\Pi_Z$ is related to the density-density correlation.
The ratio $\mtca{Z}_S/\Sigma^2$ measures therefore the fluctuation of
the quark condensate. For $N_f$ near the critical point $\ncrit$ where
$\Sigma$ vanishes, one may expect a suppression of
$\Sigma$ and an enhancement of its fluctuations $\mtca{Z}_S$.

We can express the ratio $\mtca{Z}_S/\Sigma^2$
by introducing the Gell-Mann--Oakes--Renner ratio \cite{gor},
measuring the condensate in physical units:
\begin{equation}
X_{\mtrm{GOR}}(N_f)=\frac{2m\Sigma(N_f)}{F_\pi^2M_\pi^2},
\end{equation}
where $m$ denotes the common mass of
the $N_f$ lightest quarks ($m=(m_u+m_d)/2$ for $N_f=2$).
Following the analysis of Ref.~\cite{Descotes-Girlanda-Stern}, one obtains
from Eq.~(\ref{diff}), in the approximation $Z_\mtrm{eff}\sim 2\mtca{Z}_S$:
\begin{eqnarray}
\frac{F_\pi^4\mtca{Z}_S(N_f+1)}{\Sigma^2(N_f+1)}
  &\sim& \frac{F_\pi^4Z_{\mtrm{eff}}}{2\Sigma^2(N_f+1)}\label{prerho}\\
  &=&\frac{X_{\mtrm{GOR}}(N_f)-X_{\mtrm{GOR}}(N_f+1)}
      {[X_{\mtrm{GOR}}(N_f+1)]^2}\cdot\frac{F_\pi^2}{2rM_\pi^2}+\ldots, 
      \label{ratdis}
\end{eqnarray}
where $r$ stands for $m_s/m$ and the dots denote higher-order terms.
For $N_f\ll\ncrit$, the right-hand side of Eq.~(\ref{ratdis}) is very small.
It can be illustrated by choosing $N_f=2$, $X(2)\sim 0.9$ and $r\sim 26$
(Standard $\chi$PT estimates). The difference of the
GOR ratios satisfies in this case the lower bound: $X(2)-X(3)>0.2$
\cite{paramag,Moussallam},
and we consider this bound conservatively as an equality.
In this case, the right-hand side of Eq.~(\ref{ratdis})
is of the order of $10^{-2}$ (let us notice that in this case, this quantity
is related to $16L_6(\mu)$ at a typical hadronic scale $\mu\sim M_\rho$,
c.f. Ref.~\cite{Moussallam}). The proximity of a
phase transition could be detected by a considerable increase of the ratio
(\ref{prerho}) compared to its typical size $\sim 10^{-2}$.

\subsection{Application to the lattice}

An evaluation of the inverse moments through lattice simulation 
represents a few interesting features.
We work at finite volumes: the volume dependence
is crucial to obtain information on the relevant
low-energy constants of the effective Lagrangian, and the
extrapolation to an infinite volume is avoided.
The limitation to the topologically trivial sector is natural on the lattice
by choosing strictly periodic boundary conditions. 

We do not aim at solving full QCD on the lattice. We want to compute Dirac
inverse moments, averaged over the gluonic configurations with the
statistical weight (\ref{moynu}). To perform this more limited task, we have to
know the Dirac spectrum for each gluonic configuration. It can be obtained
through the square of the Dirac operator:
${D\dirac}^2=D^2+iF^{\mu\nu}\sigma_{\mu\nu}$. It seems much simpler to
discretize this operator instead of ${D\dirac}$ itself. In particular, 
the doubling problems are not expected to arise in the spectrum of an elliptic
operator like $D^2$. It should be stressed that, while this procedure could
be applied in our particular problem, it can hardly represent a general
solution for doubling in the spectrum of lattice fermions.

For a given gluonic configuration, we can therefore compute the inverse
moments from the Dirac spectrum (which is independent of the number of
flavours). The essential contribution to each inverse moments stems
from the lowest eigenvalues. In this case, the $N_f$-dependence in the average
$\lcr\rcr\nflav_0$ is expected to arise mainly through the $N_f$-th power of
the product of the lowest Dirac eigenvalues, i.e. from the infrared part of
the truncated fermion determinant, c.f. Eq.~(\ref{truncatdet}).
The ultraviolet part of the determinant should then be included by a
matching with the
perturbative tail as discussed in Ref.~\cite{Duncan-Eichten}. A first
possibility consists in generating the gluonic configurations in the
quenched approximation, and to include explicitly
the fermion determinant in the observable.
The advantage of this method is that it would allow to change easily and
continuously $N_f$ while keeping the same set of gluonic configurations. 
On the other hand, Monte-Carlo simulation of the pure gauge theory could lead to a rather 
different distribution of small Dirac eigenvalues than a simulation including the fermion 
determinant into the statistical weight: the quenched generation of the configurations
may therefore lead to biased results, when we use these configurations
to compute quantities including explicitly a fermion determinant
as an observable. If this reweighting procedure turns out to be inefficient, 
the generation of the gluonic configurations would
have to include the product of the lowest Dirac eigenvalues in the statistical weight. The
configurations should be regenerated for each value of $N_f$.

The computation of the ratio
$\rho$ seems particularly attractive on the lattice. We have to compare five
different lattice sizes to calculate this ratio, invariant under the QCD
renormalization group and protected from \nlo\ effects. 
When $N_f$ increases, an enhancement of $1/\rho$ would clearly indicate the vicinity
of the critical point $\ncrit$ where the condensate vanishes.

\section{Conclusion}

Two descriptions of Euclidean QCD on a torus can be fruitfully matched : the
first involves the spectrum of the Dirac operator whereas the second relies
on the effective theory of Goldstone bosons.
The spontaneous breakdown of chiral symmetry can be related to
the large-volume behaviour of inverse moments of the Dirac eigenvalues,
$\sum_{n>0} 1/\lambda^k_n$, averaged over topological sets of gluonic
configurations. Because of their sensitivity to $N_f$, these inverse moments
can be used to detect chiral phase transitions occurring when the number
of massless flavours increases.

The quark condensate $\Sigma(N_f)$ is the chiral order parameter that is the
most sensitive to $N_f$. It is conceivable that just above the first
critical point $\ncrit$ where $\Sigma$ vanishes, the chiral symmetry is
only partially restored. Below this critical point, the large-volume behaviour of the inverse
spectral moments is given by the Leutwyler-Smilga sum rules
(\ref{stdls1})-(\ref{stdls2}) and it is driven by the quark
condensate (this behaviour corresponds to eigenvalues accumulating as $1/L^4$).
Above $\ncrit$, the asymptotic volume dependence of the inverse moments
changes (see Eqs.~(\ref{ext1})-(\ref{ext2})), 
corresponding to eigenvalues behaving as $1/L^2$. In this case,
the dominant contribution comes from terms in the effective Lagrangian quadratic 
in quark masses.

When $N_f$ increases and approaches $\ncrit$, the quark
condensate becomes small, and its fluctuations (related to the Zweig-rule violation in
the scalar channel) are expected to become large: the terms of the effective
Lagrangian linear and quadratic in the quark masses
may therefore contribute with a comparable magnitude.
Hence, it may become necessary to include both of them into the leading
order of the expansion of $\mtca{L}_\mtrm{eff}$, in order to derive the
large-volume behaviour of the inverse spectral moments, which interpolates between both phases.
The resulting sum rules have been analyzed in the topologically trivial sector 
$\nu=0$ (see Eqs.~(\ref{gsr22})-(\ref{endl6})). 
In particular, the formulae concerning positive inverse moments
restrict the parameters of the effective Lagrangian.

For $N_f\ll \ncrit$, the first subleading corrections to Leutwyler-Smilga
sum rules are due to
the non-zero modes, and reduce to a volume-dependent redefinition of the low-energy
constant $\Sigma$. The next-to-leading corrections to these formulae have
been calculated also for $N_f$ close to the critical point $\ncrit$.
The part arising from non-zero modes is translated into a
redefinition of the low-energy constants $\Sigma$, $\mtca{Z}_S$,
$\mtca{Z}_P$ and $\mtca{A}$. The NLO contribution
due to zero modes can be computed directly for $\nu=0$. All
NLO corrections behave as $O(L^{-2})$ relatively to the leading contribution.

We have shown that combining inverse spectral moments at different volumes
allows one to isolate the ratio of low-energy constants
$\mtca{Z}_S/\Sigma^2$ which is particularly sensitive to the chiral phase
transition. The resulting ``five-volume formula'' (\ref{latt3}) is
furthermore insensitive to NLO finite-size corrections,
and it is invariant under the QCD renormalization group.

The study of the inverse spectral moments of the Dirac operator seems a 
promising tool to investigate chiral phase transitions in
association with lattice simulations. The sums over eigenvalues can be
computed from a set of gluonic configurations with $\nu=0$
and the corresponding Dirac spectra, obtained after the diagonalization of
$D^2+F\sigma/2$. The $N_f$-dependence is explicit, via the infrared part of
the truncated fermion determinant and the finite-volume
effects are not only taken into account, but essential for our purposes.

The possibility to vary on the lattice parameters fixed in the real world, 
like $N_f$ (and $N_c$) could open a new window on the chiral
structure of QCD vacuum.
This investigation could lead to a better understanding of QCD-like theories
in general. For instance, among electroweak symmetry breaking
models, technicolor and similar proposals have often been ruled out, assuming
a smooth and simple dependence on $N_f$ and $N_c$ leading to a direct link
with actual QCD phenomenology \cite{electroweak}. If the chiral phase structure of
vector-like confining gauge theories turned out to be richer, the chiral
symmetry could be broken following a different pattern from actual QCD,
offering new possibilities for technicolor-like models \cite{phasetrans}. The
study of $N_f$-induced chiral phase transitions could therefore represent a
step towards alternative theories of electroweak symmetry breaking.

\acknowledgments % RevTex

We thank Ph.~Boucaud, L.~Girlanda, P.~Hasenfratz, H.~Leutwyler, G.~Martinelli,
B.~Moussallam and C.~Roiesnel for valuable discussions. 

\appendix

\section{Integration over unitary matrices} \label{integrunitary}

In the flavour space, we can define a complete set $\{t_a\}$ of 
$N_f^2$ Hermitian $N_f\times N_f$ matrices generating $\mtrm{U}(N_f)$.
$a$ is an index from 0 to $N_f^2-1$: $t_0$ is proportional to the
identity, and the other matrices are traceless. They are normalized by:
\begin{equation}
\langle t_at_b\rangle  = \frac{1}{2} \delta_{ab},
 \qquad \qquad \sum_a t_a t_a =\frac{1}{2} N_f,
\end{equation}
with the interesting identities for any matrices $A$ and $B$:
\begin{eqnarray}
\sum_a \langle t_a A\rangle \langle t_a B\rangle
   &=&\frac{1}{2}\langle AB\rangle,\\
\sum_a t_a A t_a &=& \frac{1}{2}\langle A\rangle, \qquad
\sum_a \langle t_a A t_a B\rangle = \frac{1}{2}\langle A\rangle
 \langle B\rangle.
\end{eqnarray}
We can decompose any complex matrix
on this basis: $X=\sum_a X_a t_a$. If we want
to perform integrations over $\mtrm{U}(N_f)$ involving a unitary matrix $U$, 
the non-vanishing integrals have as many components from $U=\sum_a U_a t_a$ as 
from $U^\dag=\sum_a U^*_a t_a$. The first ones are:
\begin{eqnarray}
\int_{\mtrm{U}(N_f)} \!\![dU]&& = 1,\\
\int_{\mtrm{U}(N_f)} \!\![dU]&& U_a U^*_b=\frac{2}{N_f} \delta_{ab},\\
\int_{\mtrm{U}(N_f)} \!\![dU]&& U_a U^*_b U_c U^*_d=
  \frac{4}{N_f^2-1} (\delta_{ab}\delta_{cd}+\delta_{ad}\delta_{bc})\\
&& -\frac{16}{N_f(N_f^2-1)}
       \langle t_a t_b t_c t_d + t_a t_d t_c t_b \rangle\nonumber,\\
\int_{\mtrm{U}(N_f)} \!\![dU]&& U_a U^*_b U_c U^*_d U_e U^*_f=
   \frac{8}{N_f(N_f^2-1)(N_f^2-4)}\\
&& \times \Big\{(N_f^2-2)
     [\delta_{ab}\delta_{cd}\delta_{ef}+\delta_{ab}\delta_{cf}\delta_{ed}
     +\delta_{ad}\delta_{cb}\delta_{ef}\nonumber\\
&& \qquad \qquad \qquad +\delta_{ad}\delta_{cf}\delta_{eb}
     +\delta_{af}\delta_{cb}\delta_{ed}+\delta_{af}\delta_{cd}\delta_{eb}]\nonumber\\
&& \quad -4N_f
     [\delta_{ab} \langle t_c t_d t_e t_f + t_c t_f t_e t_d \rangle
     +\delta_{ad} \langle t_b t_c t_f t_e + t_b t_e t_f t_c \rangle\nonumber\\
&& \qquad \qquad +\delta_{af} \langle t_b t_c t_d t_e + t_b t_e t_d t_c \rangle
     +\delta_{cb} \langle t_a t_d t_e t_f + t_a t_f t_e t_d \rangle\nonumber\\
&&\qquad \qquad +\delta_{cd} \langle t_a t_b t_e t_f + t_a t_f t_e t_b \rangle
     +\delta_{cf} \langle t_a t_b t_e t_d + t_a t_d t_e t_b \rangle\nonumber\\
&&\qquad\qquad +\delta_{eb} \langle t_a t_d t_c t_f + t_a t_f t_c t_d \rangle
     +\delta_{ed} \langle t_a t_b t_c t_f + t_a t_f t_c t_b \rangle\nonumber\\
&&\qquad \qquad+\delta_{ef} \langle t_a t_b t_c t_d + t_a t_d t_c t_b \rangle]\nonumber\\
&& \quad +16
     \langle t_a t_b t_c t_d t_e t_f +  t_a t_b t_e t_d t_c t_f 
     +  t_a t_b t_e t_f t_c t_d +  t_a t_b t_c t_f t_e t_d\nonumber\\
&&\qquad \qquad + t_a t_d t_c t_b t_e t_f +  t_a t_d t_e t_b t_c t_f
     +  t_a t_d t_e t_f t_c t_b +  t_a t_d t_c t_f t_e t_b\nonumber\\ 
&&\qquad \qquad  +  t_a t_f t_c t_d t_e t_b +  t_a t_f t_e t_d t_c t_b
     +  t_a t_f t_e t_b t_c t_d +  t_a t_f t_c t_b t_e t_d \rangle\Big\}
\nonumber.
\end{eqnarray}

\section{Leading-order generalized Lagrangian\\for two flavours}
\label{sec2flav}
For $N_f=2$, the situation is slightly different from the generic case,
because $\mtrm{SU}(2)$ representations are pseudoreal. In particular, the
correlator $\langle(\bar{u}u)(\bar{d}d)\rangle$, which defines
$\mtca{Z}_S^{(2)}$, contains a determinant-like invariant and 
is no more an order parameter.
The \lo\  of the generalized Lagrangian for $\mtrm{SU}(2)$ is \cite{twoflav}:
\begin{eqnarray}
\tilde{\mtca{L}}_2^{(2)}&=&\frac{1}{4}\left\{F^2(2)
  \langle \partial_\mu U^\dag \partial_\mu U\rangle
     -2\Sigma(2)\langle U^\dag M+M^\dag U\rangle\right.\label{gchipt2}\\
&&\quad -\mtca{A}(2)\langle (U^\dag M)^2+(M^\dag U)^2\rangle\nonumber
   -\mtca{Z}_P(2)\langle U^\dag M-M^\dag U\rangle^2\\
&&\quad \left. -\mtca{H}(2)\langle M^\dag M\rangle
      -\mtca{H}'(2)(\det M+\det M^\dag)\right\}.\nonumber
\end{eqnarray}
The new counterterm $\mtca{H}'(2)$ is consistently counted $O(p^2)$ in G$\chi$PT,
since $\det M$ involves two powers of the mass.

Despite similarities between $\mtca{H}'(2)$ and $\mtca{H}(2)$ 
(terms with no mesonic fields,
absent from the low-energy processes), $\mtca{H}'(2)$ is not necessarily
divergent.
In the Minkowskian metric, it can be defined through
the chiral limit of the Zweig-suppressed correlator:
\begin{equation}
2i\int d^4x\ e^{ipx} \langle 0 | 
   T\{\bar u u(x)\bar d d(0)\} |0\rangle 
   =\mtca{H}'^r(2)(\mu)+O(p^2)_{\mtrm{G\chi PT}}.
      \label{uudd}
\end{equation}
It is easy to prove that, in the chiral limit, the
identity operator, the quark condensate and the gluon condensate do not
contribute to the Operator Product Expansion of 
$\langle (\bar{u}u)(\bar{d}d)\rangle$~\footnote{Basically, the quark condensate
cannot appear in OPE of (\ref{uudd})
without a mass term, vanishing in the chiral limit, whereas
the discrete symmetry $u_L \to -u_L$ rules out the identity operator and the
gluon condensate. We thank B.~Moussallam for this remark.},
The correlator (\ref{uudd}) is dominated by $d=6$ operators and it behaves as
$O(1/p^4)$ for large momenta. It is therefore superconvergent.
$\mtca{H}'(2)$ can be related to the scalar spectrum through a dispersion relation
with no subtraction, similarly as in Ref.~\cite{Moussallam}. 
Despite the difficulty of estimating the
resulting integral, $\mtca{H}'(2)$ can be determined in principle from experimental
data in the $0^+$ sector, including not only the low-energy dynamics, but
also information about higher resonances. 

Since $\mtca{H}'(2)$ is free of ultraviolet divergences,
we can formally rewrite the G$\chi$PT leading order of 
the two-flavour Lagrangian in the generic
form ($N_f \geq 3$). We use the identity, true for any $2\times 2$
matrix $C$: $\langle C\rangle^2-\langle C^2\rangle =2\det C$. This leads to 
a formal identification:
\begin{eqnarray}
\mtca{A}(N_f)\leftrightarrow \mtca{A}(2)-\mtca{H}'(2)/2, &\qquad& 
  \mtca{Z}_S(N_f)\leftrightarrow \mtca{H}'(2)/4, \\
  \mtca{Z}_P(N_f)\leftrightarrow \mtca{Z}_P{(2)}+\mtca{H}'(2)/4, &\qquad&
  \mtca{H}(N_f) \leftrightarrow \mtca{H}(2),
\end{eqnarray}
which enables us to treat the two-flavour Lagrangian in the same framework
as the generic case, even though the phenomenological interpretation of 
its parameters is different.

\section{Expansion coefficients of the partition function} \label{appcoef}

This section is devoted to the calculation of the coefficients arising when
the partition function is expanded in powers of $X=ML^2$ for $N_f$ near (but
under) the critical point $\ncrit$. The main lines of the
computation are exposed in Sec.~\ref{sectechnical}, but its technical details
and the results for an arbitrary winding number are presented here. The
coefficients $\alpha_\nu$, $\beta_\nu$\ldots are defined in Eq.~(\ref{devinu}).

\subsection{Leading coefficient $\alpha_\nu$} \label{appcoefa}

To compute $\alpha_\nu(b,z,a)$, we begin with $\alpha_\nu(b,0,a)$,
given by the \lo\  in $x$ of the group integral:
\begin{equation}
I^\alpha_\nu=I_\nu(b,a;x \cdot 1)
  =\int_{\mtrm{U}(N_f)} \ [dU] (\det U)^\nu 
       \exp[bx\langle U^\dag\rangle+ax^2\langle {U^\dag}^2\rangle].
\end{equation}

We can use Weyl's formula to transform the group integral into an 
integration 
over the eigenvalues of $U$: $\exp(i\phi_k)$ ($k=1\ldots N_f$):
\begin{equation}
\int_{\mtrm{U}(N_f)} \ [dU]
  \to \frac{1}{N_f!}\int \left(\prod_{k=1}^{N_f}\frac{d\phi_k}{2\pi}\right)
   |P|^2,
\end{equation}
with $P=\prod_{k<l} \left(e^{i\phi_k}-e^{i\phi_l}\right)$.
$P$ is a linear combination of $\exp(i\sum n_k \phi_k)$, with $n_k$
integers, antisymmetric under the exchange of two angles, so that for
$k\neq l$, $n_k$ and $n_l$ must be different. Their set forms one of
the $N_f!$
permutations of $(0,1,2\ldots N_f-1)$, and $P$ collects all of them,
with a sign depending on the signature of the permutation. If the integrand
is symmetric under the angle permutations, $PP^*$ can be rewritten
\cite{Leut-Smil}:
\begin{equation}
|P|^2=N_f!\sum_{\sigma\in\mtca{P}(N_f)}
         \epsilon(\sigma) \exp\left[i\sum_{k=1}
	  ^{N_f}(\sigma(k)-k)\phi_k\right], \label{weyl}
\end{equation}
with $\mtca{P}(N_f)$ the set of the permutations over $(1\ldots N_f)$
and $\epsilon$ the signature.

The group integral $I^\alpha_\nu$ becomes:
\begin{equation}
I^\alpha_\nu = 
   \frac{1}{N_f!}\int \left(\prod_{k=1}^{N_f}\frac{d\phi_k}{2\pi}\right) |P|^2
      \prod_{k=1}^{N_f}
         \left(e^{i\nu\phi_k} 
	   \exp\left[bx e^{-i\phi_k}+ax^2 e^{-2i\phi_k}\right] \right).
	   \label{inuangle}
\end{equation}
When $|P|^2$ is replaced by its symmetrized value (\ref{weyl}),
the integrals over the angles become independent of each other: 
\begin{eqnarray}
I^\alpha_\nu &=& 
   \sum_{\sigma\in\mtca{P}(N_f)} \epsilon(\sigma)
   \prod_{k=1}^{N_f} \int \frac{d\phi_k}{2\pi}
         e^{i(k-\sigma(k)+\nu)\phi_k} 
	   \exp\left[bx e^{-i\phi_k}+ax^2 e^{-2i\phi_k}\right]\\
 &=& \sum_{\sigma\in\mtca{P}(N_f)} \epsilon(\sigma)
       \prod_{k=1}^{N_f} 
       x^{s(k)} 
     \sum_{\scriptstyle p_k+2q_k=s(k)\atop \scriptstyle p_k, q_k \geq 0} 
        \frac{1}{p_k!q_k!} b^{p_k} a^{q_k},
\end{eqnarray}
with $s(k)=k-\sigma(k)+\nu$. Obviously, if $s(k)<0$ for at least one $k$,
the permutation does not contribute. But $\mtca{P}(N_f)$
includes the identical permutation and $\nu\geq 0$: there is at least one
contributing term in $I^\alpha_\nu$, and all these contributions lead actually
to the same leading power in $x$:
\begin{equation}
  \prod_{k=1}^{N_f} x^{s(k)}=x^{\sum k-\sigma(k)+\nu}=x^{\nu N_f},
\end{equation}
which is consistent with the factor $(\det X)^\nu$ in the expansion 
(\ref{devinu}). We get therefore:
\begin{equation}
\alpha_\nu(b,z=0,a)=\sum_{m=0\ldots \nu N_f/2} b^{\nu N_f-2m} a^m c_m,
\end{equation}
with the purely combinatorial coefficients:
\begin{equation}
c_m=\sum_{\sigma\in\mtca{P}(N_f)} \epsilon(\sigma)
 \sum_{\scriptstyle  \{q_k=1\ldots s(k)/2\} \atop 
       \scriptstyle \sum q_k=m
    }
 \left[\prod_k q_k! (s(k)-2q_k)! \right]^{-1}.
\end{equation}
Another way to describe $c_m$ is the generating polynomial:
\begin{equation}
\sum_{m=0\ldots \nu N_f/2} w^m c_m
   =\left| 
   \begin{array} {ccccc}
   X_\nu & X_{\nu+1} & X_{\nu+2} & \cdots & X_{\nu+N-1}\\
   X_{\nu-1} & X_\nu & X_{\nu+1} & \cdots & X_{\nu+N-2}\\
   X_{\nu-2} & X_{\nu-1} & X_\nu & \cdots & X_{\nu+N-3}\\
   \vdots & \vdots & \vdots & \ddots & \vdots \\
   X_{\nu-N+1} & X_{\nu-N+2} & X_{\nu-N+3} & \cdots & X_\nu
   \end{array}   \right|,
\end{equation}
with the polynomials in $w$:
\begin{equation}
X_j=\sum_{q=0\ldots j/2} \frac{w^q}{q!(j-2q)!}.
\end{equation}

Since the derivatives of $I_\nu$ 
with respect to $b$ and $z$ are not independent,
Eq.~(\ref{derivinu}) yields the general expression of $\alpha_\nu$:
\begin{equation}
\alpha_\nu(b,z,a)=
   \sum_{l+2m+2p=\nu N_f} b^l a^m z^p \frac{(l+2p)!}{l!p!} c_m.
\end{equation}

\subsection{Subleading coefficients $\beta_\nu$, $\gamma_\nu$, $\delta_\nu$}
 \label{appcoefb}

We denote the various derivatives of $\alpha_\nu$:
\begin{equation}
\alpha'_\nu=\frac{\partial \alpha_\nu}{\partial b}\qquad
\dot{\alpha}_\nu=\frac{\partial \alpha_\nu}{\partial a}\qquad
\alpha''_\nu=\frac{\partial^2 \alpha_\nu}{\partial b^2}
       =\frac{\partial \alpha_\nu}{\partial z}
\end{equation}

For $b=\bar{b}$, $a=\bar{a}$, $z=\bar{z}$, for $N$ flavours, and denoting 
$K=N+|\nu|$, the coefficients are:
\begin{equation}
\beta=\alpha\frac{1}{K}(y+b^2)+\alpha'\frac{1}{KN}b
   \left(2Nz+2a+Ny\right)
           +\alpha'' \frac{2}{NK}y(Nz+a)
\end{equation}

\begin{eqnarray}
\gamma&=&
\alpha\left\{\frac{1}{K^2-1}[\frac{b^4}{2}+2b^2y+2b^2z+y^2+2z^2+2a^2]
     \right.\\
 &&\quad\left.
      -\frac{1}{K(K^2-1)}2a[b^2+2z]+\frac{(K-N)(KN+1)}{K(K^2-1)(N^2-1)}2a^2
            \right\}\nonumber\\
&&+\alpha'\left\{\frac{1}{K^2-1}b[b^2y+2b^2z+2y^2+6yz+4z^2]\right.\nonumber\\
&&\quad-\frac{1}{K(K^2-1)}2ab[y+2z]+\frac{1}{N(K^2-1)}2ab[b^2+2y+2z]\nonumber\\
&&\quad\left.
  -\frac{1}{KN(K^2-1)}2a^2b-\frac{K+N}{K(K^2-1)(N^2-1)}2a^2b\right\}
    \nonumber\\
&&+\alpha''\left\{\frac{1}{K^2-1}
        [\frac{1}{2}b^2y^2+4b^2yz+2b^2z^2+5y^2z+4z^3]
  \right.\nonumber\\
&&\quad-\frac{1}{NK(K^2-1)}8a^2z
   +\frac{1}{N(K^2-1)}4a[b^2y+b^2z+y^2+2z^2]\nonumber\\
&&\quad\left.-\frac{1}{K(K^2-1)}a[y^2+4z^2]
   +\frac{KN+1}{NK(N^2-1)(K^2-1)}2a^2b^2\right\}\nonumber\\
&&+\alpha'''\left\{\frac{1}{K^2-1}2byz[2z+y]
       +\frac{1}{N(K^2-1)}2aby[4z+y]\right.\nonumber\\
&&\quad\left.+\frac{KN+1}{NK(K^2-1)(N^2-1)}4a^2by
       \right\}\nonumber\\
&&+\alpha''''\left\{\frac{1}{K^2-1}2y^2z^2
       +\frac{1}{N(K^2-1)}4ay^2z\right.\nonumber\\
&&\quad\left.+\frac{KN+1}{NK(K^2-1)(N^2-1)}2a^2y^2
       \right\}\nonumber\\
&&-\frac{K+N}{NK(K^2-1)(N^2-1)}
    \left[\dot{\alpha} 2a^2(2z+b^2) +\dot{\alpha}' 4a^2by
          +\dot{\alpha}'' 2a^2y^2 \right]\nonumber
\end{eqnarray}

\begin{eqnarray}
\delta&=&
\alpha\left\{-\frac{1}{K(K^2-1)}
      [\frac{b^4}{2}+2b^2z+2b^2y+y^2+2z^2+2a^2]\right.\\
 &&\quad\left.
      +\frac{1}{K^2-1}2a[b^2+2z]-\frac{(K-N)(K+N)}{K(K^2-1)(N^2-1)}2a^2
            \right\}\nonumber\\
&&+\alpha'\left\{-\frac{1}{K(K^2-1)}b[b^2y+2b^2z+2y^2+6yz+4z^2]\right.
     \nonumber\\
&&\quad+\frac{1}{K^2-1}2ab[y+2z]-\frac{1}{NK(K^2-1)}2ab[2y+2z+b^2]\nonumber\\
&&\quad\left.
      +\frac{1}{N(K^2-1)}2a^2b+\frac{KN+1}{K(K^2-1)(N^2-1)}2a^2b\right\}
         \nonumber\\
&&+\alpha''\left\{-\frac{1}{K(K^2-1)}
       [\frac{1}{2}b^2y^2+4b^2yz+2b^2z^2+5y^2z+4z^3]
  \right.\nonumber\\
&&\quad+\frac{1}{N(K^2-1)}8a^2z
   -\frac{1}{NK(K^2-1)}4a[b^2y+b^2z+y^2+2z^2]\nonumber\\
&&\quad\left.+\frac{1}{K^2-1}a[y^2+4z^2]
   -\frac{K+N}{NK(N^2-1)(K^2-1)}2a^2b^2\right\}\nonumber\\
&&+\alpha'''\left\{-\frac{1}{K(K^2-1)}2byz[2z+y]
       -\frac{1}{NK(K^2-1)}2aby[4z+y]\right.\nonumber\\
&&\quad\left.-\frac{K+N}{NK(K^2-1)(N^2-1)}4a^2by
       \right\}\nonumber\\
&&+\alpha''''\left\{-\frac{1}{K(K^2-1)}2y^2z^2
       -\frac{1}{NK(K^2-1)}4ay^2z\right.\nonumber\\
&&\quad\left.-\frac{K+N}{NK(K^2-1)(N^2-1)}2a^2y^2
       \right\}\nonumber\\
&&+\frac{KN+1}{NK(K^2-1)(N^2-1)}
    \left[\dot{\alpha} 2a^2(2z+b^2) +\dot{\alpha}' 4a^2by
          +\dot{\alpha}'' 2a^2y^2 \right]\nonumber
\end{eqnarray}

\section{Dimensional regularization on a torus} \label{appdimreg}

Following the regularization procedure described by Hasenfratz and
Leut\-wy\-ler \cite{Hasen-Leut}, we want to regularize sums like:
\begin{equation}
G_H=\frac{1}{V}\sum_p H(p),
\end{equation}
where $H$ is a function and $p$ is summed over $2\pi/L \cdot {\mathbb Z}^4$.
The Fourier transform of $H(p)$ is:
\begin{equation}
\tilde{H}(x)=\int \frac{d^dp}{(2\pi)^d} e^{ipx} H(p),
\end{equation}
and fulfills the identity:
\begin{equation}
G_H=\frac{1}{V} \sum_p H(p) = \sum_l \tilde{H}(l),
\end{equation}
where $l$ is summed over $L\cdot Z^4$.
Because of the relation:
\begin{equation}
\lim_{V\to\infty} G_H= \lim_{V\to\infty}\frac{1}{V} \sum_p H(p) =
   \int \frac{d^dp}{(2\pi)^d} H(p)=\tilde{H}(0),
\end{equation}
it is possible to separate in $G_H$ the cut-off and the volume dependences:
\begin{equation}
G_H=\lim_{V\to\infty} G_H + g_H, \qquad \qquad g_H = {\sum_l}' \tilde{H}(l).
  \label{separdepend}
\end{equation}
The infinite-volume limit of $G_H$ contains the divergences for $d\to 4$ and
has to be regularized, whereas $g_H$ depends only on the volume.

For $v=\sum'1/n^2$, we have the relations:
\begin{equation}
\frac{1}{V} {\sum_p}' \frac{1}{p^2}
   =\lim_{M\to 0} \left[\frac{1}{V} \sum_p \frac{1}{p^2+M^2}
        -\frac{1}{VM^2}\right]
   =\lim_{M\to 0} \left[G_H-\frac{1}{VM^2}\right].
\end{equation}
with $H(p)=1/(p^2+M^2)$. In the case of the dimensional regularization, 
(\ref{separdepend}) involves:
\begin{equation}
G_H=\frac{M^2}{8\pi^2}(\ln M+c_1)+g_H, \qquad 
g_H=\frac{1}{VM^2}-\frac{\beta_1}{L^2}+O(M^2),
\end{equation}
where $c_1$ contains a pole for $d=4$, and $\beta_1$ is a constant called
``shape coefficient'', depending on the geometry of the box. For a
four-dimensional torus, $\beta_1=0.1405$. The dimensional regularization
yields finally:
\begin{equation}
v=\sum_{n\neq 0} \frac{1}{n^2}\leftrightarrow-4\pi^2\beta_1
\end{equation}

For $u$, we can follow the same guideline and take $H(p)=1$ for $p\neq 0$
and $H(0)=0$. Its Fourier transform is $\tilde{H}(l)=\delta^{(4)}(l)$.
$g_H$ vanishes, and we know that dimensionally regularized integrals like
$\int \frac{d^dp}{(2\pi)^d}$
vanish as well, so that:
\begin{equation}
u=\sum_{n\neq 0} 1\leftrightarrow 0.
\end{equation}

\begin{thebibliography}{99}

\bibitem{daphne}
L.~Maiani, G.~Pancheri and N.~Paver Eds.,
\emph{The second DAPHNE physics handbook},
Frascati, Italy: INFN (1995).

\bibitem{mainz}
A.M.~Bernstein, D.~Drechsel and T.~Walcher Eds.,
\emph{Chiral dynamics: Theory and experiment.  Proceedings, Workshop, Mainz,
Germany, September 1-5, 1997}, Berlin, Germany: Springer (1998).

\bibitem{exper}
M.~Baillargeon and P.J.~Franzini, in Ref.~\cite{daphne},
hep-ph/9407277. 
%%CITATION = HEP-PH 9407277;%%

J.~Lowe in Ref.~\cite{mainz}, J.~Schacher in \emph{id.}, hep-ph/9711361.
%%CITATION = HEP-PH 9711361;%%

\bibitem{betafunc}
D.J.~Gross and F.~Wilczek,
%`Ultraviolet Behavior Of Nonabelian Gauge Theories,''
Phys.\ Rev.\ Lett.\ {\bf 30} (1973) 1343.
%%CITATION = PRLTA,30,1343;%%

\bibitem{Banks-Zaks}
T.~Banks and A.~Zaks,
%`On The Phase Structure Of Vector - Like Gauge Theories With Massless
% Fermions,''
Nucl.\ Phys.\ {\bf B196} (1982) 189.
%%CITATION = NUPHA,B196,189;%%

\bibitem{Grunberg}
E.~Gardi and G.~Grunberg,
%`The conformal window in QCD and supersymmetric QCD,''
J.\ High\ Energy\ Phys.\ {\bf 03} (1999) 024, hep-th/9810192.
%%CITATION = JHEPA,9903,024;%%

\bibitem{gapeq}
T.~Appelquist, A.~Ratnaweera, J.~Terning and L.C.~Wijewardhana,
%`The phase structure of an SU(N) gauge theory with N(f) flavors,''
Phys.\ Rev.{\bf\ D58} (1998) 105017, hep-ph/9806472.
%%CITATION = PHRVA,D58,105017;%%

\bibitem{instmodel}
M.~Velkovsky and E.~Shuryak,
%`QCD with large number of quarks: Effects of the instanton  anti-instanton
% pairs,''
Phys.\ Lett.{\bf\ B437} (1998) 398, hep-ph/9703345.
%%CITATION = PHLTA,B437,398;%%

T.~Appelquist and S.B.~Selipsky,
%`Instantons and the chiral phase transition,''
Phys.\ Lett.{\bf\ B400} (1997) 364, hep-ph/9702404.
%%CITATION = PHLTA,B400,364;%%

\bibitem{Descotes-Girlanda-Stern}
S.~Descotes, L.~Girlanda and J.~Stern,
%`Paramagnetic effect of light quark loops on Chiral Symmetry Breaking,''
hep-ph/9910537.
%%CITATION = HEP-PH 9910537;%%

\bibitem{paramag} 
J.~Schwinger,
%`The Theory Of Quantized Fields. 5,''
Phys.\ Rev.\ {\bf 93} (1954) 615.
%%CITATION = PHRVA,93,615;%%

R.~Schrader and R.~Seiler,
%`A Uniform Lower Bound On The Renormalized Scalar Euclidean Functional
% Determinant,''
Comm.\ Math.\ Phys.\ {\bf 61} (1978) 169.
%%CITATION = CMPHA,61,169;%%

H.~Hogreve, R.~Schrader and R.~Seiler,
%`A Conjecture On The Spinor Functional Determinant,''
Nucl.\ Phys.\ {\bf B142} (1978) 525.
%%CITATION = NUPHA,B142,525;%%

D.~Brydges, J.~Frohlich and E.~Seiler,
%`On The Construction Of Quantized Gauge Fields. I. General Results,''
Ann.\ Phys.\ {\bf 121} (1979) 227.
%%CITATION = APNYA,121,227;%%

J.E.~Avron and B.~Simon,
Phys.\ Lett.\ {\bf A75} (1980) 41.
%%CITATION = PHLTA,75A,41;%%


\bibitem{Vafa-Witten}
C.~Vafa and E.~Witten,
%`Eigenvalue Inequalities For Fermions In Gauge Theories,''
Commun.\ Math.\ Phys.\ {\bf 95} (1984) 257.
%%CITATION = CMPHA,95,257;%%

\bibitem{Banks-Casher} 
T.~Banks and A.~Casher,
%`Chiral Symmetry Breaking In Confining Theories,''
Nucl.\ Phys.\ {\bf B169} (1980) 103.
%%CITATION = NUPHA,B169,103;%%

\bibitem{Stern} 
J.~Stern,
%`Two alternatives of spontaneous chiral symmetry breaking in QCD,''
hep-ph/9801282.
%%CITATION = HEP-PH 9801282;%%

\bibitem{randmat}
J.J.~Verbaarschot,
%`The infrared limit of the QCD Dirac spectrum and applications of chiral
% random matrix theory to QCD,''
hep-ph/9902394.
%%CITATION = HEP-PH 9902394;%%

\bibitem{instantons}
T.~Schafer and E.V.~Shuryak,
%`Instantons in QCD,''
Rev.\ Mod.\ Phys.\ {\bf 70} (1998) 323,
hep-ph/9610451.
%%CITATION = RMPHA,70,323;%%

\bibitem{scalar}
S.~Spanier and N.~Tornqvist in
%`Note on scalar mesons: in Review of Particle Physics (RPP 1998),''
Eur.\ Phys.\ J.\ {\bf C3} (1998) 390.
%%CITATION = EPHJA,C3,390;%%

For a recent discussion, M.R.~Pennington,
%`Riddle of the scalars: Where is the sigma?,''
hep-ph/9905241, and references therein.
%%CITATION = HEP-PH 9905241;%%

\bibitem{Shifman}
I.I.~Kogan, A.~Kovner and M.~Shifman,
%`Chiral symmetry breaking without bilinear condensates, unbroken axial
% Z(N) symmetry, and exact QCD inequalities,''
Phys.\ Rev.\ {\bf\ D59} (1999) 016001, hep-ph/9807286.
%%CITATION = PHRVA,D59,016001;%%

\bibitem{lattflav1} 
Y.~Iwasaki, K.~Kanaya, S.~Kaya, S.~Sakai and T.~Yoshie,
%`Quantum chromodynamics with many flavors,''
Prog.\ Theor.\ Phys.\ Suppl.\ {\bf 131} (1998) 415,
hep-lat/9804005.
%%CITATION = PTPSA,131,415;%%

\bibitem{lattflav2} 
D.~Chen and R.D.~Mawhinney,
%`Dependence of QCD hadron masses on the number of dynamical quarks,''
Nucl.\ Phys.\ Proc.\ Suppl.\ {\bf 53} (1997) 216,
hep-lat/9705029.
%%CITATION = NUPHZ,53,216;%%

R.D.~Mawhinney,
%`Evidence for pronounced quark loop effects in QCD,''
Nucl.\ Phys.\ Proc.\ Suppl.\ {\bf 60A} (1998) 306,
hep-lat/9705031.
%%CITATION = NUPHZ,60A,306;%%

C.~Sui,
%QCD with zero, two and four flavors of light quarks: Results from  QCDSP,''
Nucl.\ Phys.\ Proc.\ Suppl.\ {\bf 73} (1999) 228, hep-lat/9811011.
%%CITATION = NUPHZ,73,228;%%

\bibitem{Moussallam}
B.~Moussallam,
%N_f Dependence of the Quark Condensate from a Chiral Sum Rule,''
Eur.\ Phys.\ J.\  {\bf C14} (2000) 111,
hep-ph/9909292.
%%CITATION = HEP-PH 9909292;%%

B.~Moussallam,
%`Flavor stability of the chiral vacuum and scalar meson dynamics,''
hep-ph/0005245.
%%CITATION = HEP-PH 0005245;%%

S.~Descotes and J.~Stern,
%`Vacuum fluctuations of \bar{q}q and values of low-energy constants,''
hep-ph/0007082.
%%CITATION = HEP-PH 0007082;%%

\bibitem{Leut-Smil}
H.~Leutwyler and A.~Smilga,
%Spectrum of Dirac operator and role of winding number in QCD,''
Phys.\ Rev.\ {\bf D46} (1992) 5607.
%%CITATION = PHRVA,D46,5607;%%

\bibitem{schipt} 
J.~Gasser and H.~Leutwyler,
%Chiral Perturbation Theory To One Loop,''
Ann.\ Phys.\ {\bf 158} (1984) 142 ;
%%CITATION = APNYA,158,142;%%
%Chiral Perturbation Theory: Expansions In The Mass Of The Strange Quark,''
Nucl.\ Phys.\ {\bf B250} (1985) 465.
%%CITATION = NUPHA,B250,465;%%

\bibitem{gchipt} 
N.H.~Fuchs, H.~Sazdjian and J.~Stern,
%How to probe the scale of (anti-q q) in chiral perturbation theory,''
Phys.\ Lett.\ {\bf B269} (1991) 183;
%%CITATION = PHLTA,B269,183;%%
%What pi - pi scattering tells us about chiral perturbation theory,''
Phys.\ Rev.\ {\bf D47} (1993) 3814.
%%CITATION = PHRVA,D47,3814;%%

M.~Knecht and J.~Stern, in Ref.~\cite{daphne}, hep-ph/9411253.
%Generalized chiral perturbation theory,''
%%CITATION = HEP-PH 9411253;%%

J.~Stern,
%Light quark masses and condensates in QCD,''
hep-ph/9712438.
%%CITATION = HEP-PH 9712438;%%

M.~Knecht, B.~Moussallam, J.~Stern and N.H.~Fuchs,
%The Low-energy pi pi amplitude to one and two loops,''
Nucl.\ Phys.\ {\bf B457} (1995) 513.
%%CITATION = NUPHA,B457,513;%%

\bibitem{Gass-Leut2} 
J.~Gasser and H.~Leutwyler,
%`Spontaneously Broken Symmetries: Effective Lagrangians At Finite Volume,''
Nucl.\ Phys.\ {\bf B307} (1988) 763.
%%CITATION = NUPHA,B307,763;%%

\bibitem{thetavac} 
C.G.~Callan, R.F.~Dashen and D.J.~Gross,
%`The structure of the gauge theory vacuum,''
Phys.\ Lett.\ {\bf 63B} (1976) 334.
%%CITATION = PHLTA,63B,334;%%

R.~Jackiw and C.~Rebbi,
%`Vacuum periodicity in a Yang-Mills quantum theory,''
Phys.\ Rev.\ Lett.\ {\bf 37} (1976) 172.
%%CITATION = PRLTA,37,172;%%

\bibitem{`t Hooft}
G.~'t Hooft,
%`Lattice regularization of gauge theories without loss of chiral symmetry,''
Phys.\ Lett.\ {\bf B349}, 491 (1995),
hep-th/9411228.
%%CITATION = PHLTA,B349,491;%%

\bibitem{Gass-Leut3} 
J.~Gasser and H.~Leutwyler,
%`Light Quarks At Low Temperatures,''
Phys.\ Lett.\ {\bf 184B} (1987) 83;
%%CITATION = PHLTA,184B,83;%%
%`Thermodynamics Of Chiral Symmetry,''
Phys.\ Lett.\ {\bf 188B} (1987) 477.
%%CITATION = PHLTA,188B,477;%%

\bibitem{itzyk-zuber}
A.~D.~Jackson, M.~K.~Sener and J.~J.~Verbaarschot,
%`Finite volume partition functions and Itzykson-Zuber integrals,''
Phys.\ Lett.\  {\bf B387} (1996) 355,
hep-th/9605183.
%%CITATION = HEP-TH 9605183;%%

\bibitem{Bijnens}
J.~Bijnens, G.~Ecker and J.~Gasser in Ref.~\cite{daphne},
%`Chiral perturbation theory,''
hep-ph/9411232.
%%CITATION = HEP-PH 9411232;%%

\bibitem{Hasen-Leut} 
P.~Hasenfratz and H.~Leutwyler,
%`Goldstone Boson Related Finite Size Effects In Field Theory And Critical
% Phenomena With O(N) Symmetry,''
Nucl.\ Phys.\ {\bf B343} (1990) 241.
%%CITATION = NUPHA,B343,241;%%
 
\bibitem{gor}
M.~Gell-Mann, R.J.~Oakes and B.~Renner,
%`Behavior Of Current Divergences Under SU(3) X SU(3),''
Phys. Rev. {\bf 175} (1968) 2195.
%%CITATION = PHRVA,175,2195;%%

\bibitem{Duncan-Eichten} 
A.~Duncan, E.~Eichten and H.~Thacker,
%`An efficient algorithm for QCD with light dynamical quarks,''
Phys.\ Rev.\ {\bf D59} (1999) 014505,
hep-lat/9806020;
%%CITATION = PHRVA,D59,014505;%%
%`Truncated determinant approach to light dynamical quarks,''
Nucl.\ Phys.\ Proc.\ Suppl.\ {\bf 73} (1999) 837,
hep-lat/9809144 and
%%CITATION = NUPHZ,73,837;%%
%`Matching the high momentum modes in a truncated determinant algorithm,''
840,
hep-lat/9809117.
%%CITATION = NUPHZ,73,840;%%

A.~Duncan, E.~Eichten, R.~Roskies and H.~Thacker,
%`Loop representations of the quark determinant in lattice QCD,''
Phys.\ Rev.\ {\bf D60} (1999) 054505,
hep-lat/9902015.
%%CITATION = PHRVA,D60,054505;%%

\bibitem{electroweak}
E.~Farhi and L.~Susskind,
%`Technicolor,''
Phys.\ Rept.\ {\bf 74} (1981) 277.
%%CITATION = PRPLC,74,277;%%

M.E.~Peskin and T.~Takeuchi,
%`Estimation of oblique electroweak corrections,''
Phys.\ Rev.\ {\bf D46} (1992) 381.
%%CITATION = PHRVA,D46,381;%%

\bibitem{phasetrans}
T.~Appelquist, P.S.~Rodrigues da Silva and F.~Sannino,
%`Enhanced global symmetries and the chiral phase transition,''
hep-ph/9906555.
%%CITATION = HEP-PH 9906555;%%

T.~Appelquist, J.~Terning and L.C.~Wijewardhana,
%`Postmodern technicolor,''
Phys.\ Rev.\ Lett.\ {\bf 79} (1997) 2767,
hep-ph/9706238.
%%CITATION = PRLTA,79,2767;%%

\bibitem{twoflav}
L.~Girlanda and J.~Stern,
%`The decay tau --> 3pi nu/tau as a probe of the mechanism of dynamical
% chiral symmetry breaking,''
hep-ph/9906489.
%%CITATION = HEP-PH 9906489;%%

L.~Ametller, J.~Kambor, M.~Knecht and P.~Talavera,
%`Low-energy photon photon fusion into three pions in generalized chiral
% perturbation theory,''
Phys.\ Rev.\ {\bf D60} (1999) 094003,
hep-ph/9904452.
%%CITATION = PHRVA,D60,094003;%%

L.~Girlanda, PhD Thesis, Universit\'e Paris-XI, Orsay, France (1999).

\end{thebibliography}
\end{document}